%% file: arxiv_version.tex
\documentclass[12pt]{article}
\input{preamble}

\title{Improving on a Lottery: Efficient Estimation of Optimal Assignment Rules\thanks{Alphabetical ordering of authors: all authors contributed equally to this work.}}
\author{Yue Fang\thanks{The Chinese University of Hong Kong, Shenzhen. Email: \url{fangyue@cuhk.edu.cn}. Fang gratefully acknowledges the financial support from
the National Natural Science Foundation of China (No.
72503208).} \and Geert Ridder\thanks{University of Southern California. Email: \url{ridder@usc.edu}.} \and Haitian Xie\thanks{Peking University. Email: \url{xht@gsm.pku.edu.cn}. Xie gratefully acknowledges the financial support from the National Natural Science Foundation of China (No. 72403008, No. 72495123).}}
\date{\today}

\begin{document}

\maketitle

\begin{abstract}
Scarce opportunities are often allocated by lotteries. We study how to improve such allocations by estimating optimal assignment rules that maximize welfare net of a Kullback--Leibler penalty for departing from the benchmark randomization. The framework covers discrete, continuous, and mixed treatments. Regret is asymptotically quadratic in the estimation error, so inefficient estimation raises the mean of limiting regret, not merely its dispersion. We show that inverse probability weighting with known assignment probabilities is inefficient, whereas estimated-propensity and doubly robust welfare criteria attain the efficient regret distribution. Simulations and a commitment-savings application quantify the resulting precision gains.

\end{abstract}
\newpage

\section{Introduction}  

Many scarce opportunities are allocated by lottery, sometimes with weights reflecting administrative priorities. Examples include H-1B visas \citep{pathak2025immigration}, automobile licenses \citep{li2018better}, Medicaid coverage in Oregon \citep{finkelstein2012oregon}, and seats in oversubscribed charter schools \citep{abdulkadirouglu2011accountability}. Such mechanisms arise when resources must be rationed, claims are only partly comparable, prices are infeasible or undesirable, and the procedural fairness or legitimacy of the allocation process matters. A related market-design literature studies randomized rules for assigning positions, houses, and courses \citep{hylland1979efficient,abdulkadiroglu1998random,budish2013designing}. Random assignment also plays a central role when the objective is learning rather than immediate allocation: experiments randomize access to treatments such as training slots and new savings products to identify their effects and guide future policy, as in the settings underlying our simulations and empirical application. Adaptive experiments make this link between allocation and learning explicit: assignment probabilities are updated across waves, tilted toward better-performing treatments while remaining randomized so that learning can continue \citep{kasy2021adaptive}.

Data from such randomized assignment can be used to estimate improved assignment rules. A standard empirical approach estimates the expected counterfactual welfare generated by each candidate rule and then chooses the welfare-maximizing rule within a prespecified class. Under unpenalized outcome-welfare maximization, however, the optimal rule is often deterministic: individuals with sufficiently high predicted gains receive the treatment, while others do not. This logic underlies empirical welfare maximization and related treatment-choice methods; see, among others, \citet{kitagawa2018should}, \citet{athey2021policy}, and \citet{mbakop2021model}. These methods are well suited to settings in which the planner can redesign assignment from a blank slate. In the lottery settings above, by contrast, replacing an existing randomized mechanism with a hard deterministic rule may discard precisely the procedural or informational features that made randomization attractive in the first place.

We take a different starting point by treating the existing randomized rule as the benchmark. Our contribution is not to solve the underlying mechanism-design problem: the benchmark is given. It may be a status-quo lottery, a priority-weighted administrative rule, or the randomized assignment rule used in an experiment. We do not require it to satisfy a universal fairness axiom. Rather, we interpret it as the institutionally accepted assignment rule whose procedural features the planner may wish to preserve while improving outcomes. The planner chooses a randomized assignment rule from a specified class and evaluates it using a penalized welfare criterion that combines expected counterfactual outcomes with a penalty on the rule's average Kullback--Leibler divergence from the benchmark. The empirical question is therefore not only who should receive the treatment under unrestricted outcome-welfare maximization, but how much the planner can improve outcomes by tilting assignment probabilities while maintaining controlled proximity to the benchmark.

Benchmark proximity can capture several features of the planner's problem. First, when assignment-relevant variables are reported or otherwise manipulable, a deterministic threshold creates a bright line: a small change in the reported covariate can move assignment from probability zero to probability one. A smooth randomized rule weakens this incentive by making assignment probabilities vary gradually. Randomization can therefore reduce manipulation incentives rather than merely soften a deterministic allocation. Second, the benchmark may carry legal, political, or institutional legitimacy: an institution may have committed to treating similarly situated claimants similarly, or may operate under an equal-treatment mandate. In such cases, proximity to the benchmark is a constraint on feasible reform rather than a mere taste for smoothness. Third, the penalty has a structural information-cost interpretation. When changing assignment odds is costly to implement, justify, or communicate, a Kullback--Leibler cost leads the planner to tilt benchmark probabilities toward higher-payoff alternatives rather than abandon the benchmark altogether \citep{mattsson2002probabilistic,matejka2015rational}. Fourth, continued randomization preserves overlap and supports future learning. This is especially important in adaptive experiments, where each wave's rule must stay randomized for subsequent evaluation to be possible \citep{kasy2021adaptive}. In each case, proximity to the benchmark is part of the economic problem, not merely an analyst's regularization device.

These considerations are absent from a pure outcome-welfare criterion. Because expected outcome welfare is linear in the assignment rule, its unrestricted maximizer over a rich randomized class is typically a deterministic extreme point. The divergence penalty changes the target. Once benchmark proximity is part of the planner's criterion, the optimal rule is an interior randomized rule that tilts benchmark probabilities toward higher-welfare treatments. The target throughout is the best rule in a chosen interpretable class, not the unrestricted first best. Our paper studies how to estimate this target efficiently, and how first-order inefficiency in estimating the rule translates into regret.

We make three contributions. First, we develop semiparametric efficiency theory for benchmark-penalized assignment with binary, multivalued, continuous, and mixed treatments. The target is the best rule in a specified smooth class of randomized assignment rules. The theory yields a sharp decision-theoretic implication: regret relative to the best-in-class rule, multiplied by the sample size, converges in distribution to a quadratic form in the limiting estimation error of the rule parameter. The mean of this limiting regret is determined by the covariance of that error. Thus, first-order inefficiency in estimating the rule raises expected limiting regret, not merely its dispersion. Inefficient estimation therefore has consequences beyond wider confidence intervals for a fixed parameter: it systematically worsens the welfare performance of the rule ultimately chosen.

Second, we compare common criteria for estimating the assignment rule. We show that inverse probability weighting (IPW) with the true propensity score, i.e., the observed-data assignment probabilities, known by design in an experiment, contains an orthogonal noise component that is removed by estimated-propensity IPW and by doubly robust estimation. The mechanism is the efficiency phenomenon of \citet{hirano2003efficient}, henceforth HIR: estimating a propensity score that is known by design exploits overidentifying restrictions that true-propensity weighting leaves unused \citep{chen2025local}. What is new here is not the phenomenon but its implication for assignment-rule choice. Because regret is asymptotically quadratic in the estimation error, the HIR efficiency gain does not merely reduce the variance of a fixed causal estimate; it lowers expected regret in the planner's own welfare units. The familiar efficiency comparison therefore acquires a direct decision-theoretic interpretation.

Third, we show that the same force arises outside the smooth penalized problem. For unpenalized deterministic binary assignment over a class of finite VC dimension, we compare the limiting Gaussian welfare processes that enter standard regret bounds. The process generated by true-propensity IPW decomposes exactly into the process generated by estimated-propensity IPW plus an independent centered Gaussian noise process. Hence the expectation of any convex functional of the welfare process is weakly larger under true-propensity weighting, including the expected supremum that governs the regret upper bound. The inefficiency that raises the mean of penalized regret also inflates the welfare-process bound for deterministic rules. The case for efficient criteria does not rest on the penalty. The practical message is immediate: even in a randomized experiment, where assignment probabilities are known by design, weighting by the known propensity score is not the precision benchmark.

The simulations and empirical application illustrate complementary parts of the theory. In simulations calibrated to the Job Training Partnership Act data, estimated-propensity IPW and doubly robust estimation deliver lower mean penalized-welfare regret than true-propensity IPW, as predicted by the efficiency theory. In the empirical application, we revisit the commitment-savings experiment of \citet{ashraf2006tying}. We estimate benchmark-centered softmax rules that tilt assignment probabilities toward higher expected savings, trace the welfare--divergence frontier, report the implied changes in assignment probabilities, and quantify the precision gains from efficient estimation.

\paragraph{Related literature.}
This paper contributes to the econometric literature on data-driven treatment choice. \citet{manski2004statistical} and \citet{stoye2009minimax} study treatment choice under finite-sample and minimax-regret criteria. \citet{hirano2009asymptotics} develop local asymptotic theory for statistical treatment rules and show that, when the optimal rule is a discontinuous function of the underlying parameter, treatment choice is a nonregular problem to which standard efficiency bounds do not directly apply. Our benchmark-penalized problem is different: with a fixed positive penalty, the best-in-class rule is an interior finite-dimensional target that is regular under our smoothness conditions. The convolution theorem therefore applies, and first-order semiparametric efficiency acquires the second-order regret interpretation developed in Section~\ref{sec:efficiency}. The unpenalized deterministic boundary case in Section~\ref{sec:regret_upper} retains the nonregular character emphasized by \citet{hirano2009asymptotics}.

A large methodological literature studies the estimation and choice of treatment rules. \citet{kitagawa2018should} introduce empirical welfare maximization for binary treatments, and \citet{athey2021policy} study efficient welfare estimation for treatment rules with observational data. \citet{mbakop2021model} analyze model selection for treatment choice. \citet{zhou2023offline} and \citet{fang2025model} study multivalued treatments, while \citet{kallus2018policy} and \citet{ai2026data} consider continuous-treatment assignment. Much of this literature focuses on deterministic rules under unpenalized outcome-welfare criteria and derives nonasymptotic regret guarantees. We instead study smooth randomized assignment rules under a benchmark-penalized welfare criterion, deriving semiparametric efficiency bounds and second-order regret distributions. A complementary literature designs adaptive experiments that update randomized assignment rules across waves \citep{kasy2021adaptive}. Our analysis concerns the static problem of estimating an improved randomized rule from data generated under a fixed benchmark; it does not address inference with adaptively collected data \citep{hadad2021confidence}.

The paper is also motivated by the economics of randomized allocation mechanisms. Randomization is a central design tool for allocating school seats, course seats, visas, and other indivisible scarce opportunities \citep{hylland1979efficient,abdulkadiroglu1998random,budish2013designing}. Empirical studies of actual allocation mechanisms show that the choice of lottery or rationing rule can have important welfare consequences \citep{pathak2025immigration,li2018better,finkelstein2012oregon,abdulkadirouglu2011accountability}. More broadly, urban allocation and transportation policies can also have substantial welfare and distributional effects \citep{barwick2024efficiency}.

Finally, the paper contributes to semiparametric efficiency theory for causal parameters. Classical work includes \citet{hahn1998role}, \citet{hirano2003efficient}, and \citet{chen2008semiparametric}; more general treatment-effect functionals are studied by \citet{ai2021unified}. Our estimand differs from standard average treatment effects because it is the optimizer of a counterfactual assignment criterion rather than a fixed welfare functional. As a result, the convolution theorem has a direct implication for decisions: excess variance in the first-order distribution of the assignment-rule estimator raises the mean of the second-order regret distribution.

\paragraph{Organization of the paper.}
Section~\ref{sec:model} introduces the general-treatment assignment framework and the benchmark-penalized welfare criterion. Section~\ref{sec:efficiency} develops the semiparametric efficiency theory for smooth randomized assignment rules and derives the associated penalized-welfare regret distribution. Section~\ref{sec:welfare_est} analyzes IPW and doubly robust criteria and establishes the HIR phenomenon for assignment-rule estimation. Section~\ref{sec:regret_upper} studies welfare-process bounds for unpenalized deterministic rules. Sections~\ref{sec:simulation} and~\ref{sec:empirical} present the calibrated simulations and the commitment-savings application. Section~\ref{sec:conclude} concludes.

\section{Model and Assignment Criterion}\label{sec:model}

We observe an i.i.d. sample \((Z_i)_{i=1}^n\) from the distribution of \(Z=(Y,T,X)\). The treatment variable \(T\) takes values in a Borel set \(\mathcal T\subset\mathbb R\), allowing for discrete, continuous, or mixed support. Let \(Y(t)\) denote the potential outcome under treatment level \(t\in\mathcal T\), and let the observed outcome be \(Y=Y(T)\). The covariates \(X\) take values in \(\mathcal X\subset\mathbb R^d\). Let \(F_{T| X}(\cdot| x)\) denote the conditional law of the observed treatment given covariates. This law enters identification and estimation but is not part of the planner's objective. Density notation, including the generalized propensity score, is introduced only when needed for the efficiency analysis.

Assignment rules are indexed by \(\theta\in\Theta\subset\mathbb R^p\). For each \(\theta\), a candidate assignment rule \(\Pi_\theta(\cdot| x)\) is a probability kernel from covariates to treatments: for each \(x\), \(\Pi_\theta(\cdot| x)\) is a probability distribution on \(\mathcal T\), and for each measurable set \(B\subseteq\mathcal T\), the map \((x,\theta)\mapsto \Pi_\theta(B| x)\) is measurable. This notation covers binary, multivalued, continuous, and mixed treatments. It also covers deterministic rules as degenerate kernels; these are useful for the unpenalized boundary case, whereas the fixed-penalty regularity theory focuses on smooth randomized rules that are absolutely continuous with respect to the benchmark.

We also fix a known benchmark assignment rule, denoted \(\Pi^{\mathrm b}(\cdot| x)\). The benchmark may be a status-quo lottery, a priority-weighted administrative rule, a uniform randomized rule, or the randomized assignment rule used in an experiment. The benchmark assignment rule is part of the planner's objective. It is conceptually distinct from the observed-data treatment law \(F_{T| X}\). In a randomized experiment the two may coincide numerically if the experimental design is the benchmark rule, but they play different roles: \(\Pi^{\mathrm b}\) defines the rule relative to which departures are penalized, whereas \(F_{T| X}\) determines how counterfactual welfare is identified and estimated.

For probability measures \(Q\) and \(Q^{\mathrm b}\) on \(\mathcal T\), write
\[
    \mathrm{KL}\{Q\|Q^{\mathrm b}\}
    = \int_{\mathcal T}
    \log\!\left(\frac{dQ}{dQ^{\mathrm b}}(t)\right) Q(dt),
\]
when \(Q\ll Q^{\mathrm b}\), and set the divergence equal to \(+\infty\) otherwise.

For \(\lambda\geq0\), the planner evaluates a candidate rule \(\Pi_\theta\) by
\begin{align}
    W_{P,\lambda}(\theta)
    =
    \mathbb E_P\left[
        \int_{\mathcal T} Y(t)\Pi_\theta(dt| X)
    \right]
    -
    \lambda
    \mathbb E_P\left[
        \mathrm{KL}\{\Pi_\theta(\cdot| X)\|
        \Pi^{\mathrm b}(\cdot| X)\}
    \right].
    \label{eq:criterion-potential}
\end{align}
When \(\lambda=0\), the KL term is omitted and the criterion is interpreted as the usual outcome-welfare criterion. When there is no ambiguity, we suppress the subscript \(P\) and write \(W_\lambda(\theta)\).

The first term in \eqref{eq:criterion-potential} is outcome welfare: the expected counterfactual outcome generated by the assignment rule, measured in the same units as \(Y\). The second term is the average Kullback--Leibler divergence of the candidate rule from the benchmark rule. Thus \(W_{P,\lambda}\) combines outcome improvement with a penalty for departing from the benchmark. Larger values of \(\lambda\) place more weight on proximity to \(\Pi^{\mathrm b}\), and smaller values place more weight on outcome welfare. The boundary case \(\lambda=0\) is the usual outcome-welfare criterion and is studied separately in Section~\ref{sec:regret_upper}. Sections \ref{sec:efficiency} and \ref{sec:welfare_est} focus on the benchmark-improvement problem with fixed \(\lambda>0\). Section~\ref{subsec:choosing-lambda} discusses how to interpret and report \(\lambda\) empirically.

The identification of \eqref{eq:criterion-potential} relies on the following conditions.

\begin{assumption}\label{ass:unconfoundedness}
For each \(t\in\mathcal T\), \(Y(t)\perp T| X\).
\end{assumption}

\begin{assumption}\label{ass:common-support}
There exists a set \(\mathcal X_0\subseteq\mathcal X\) with
\(P_X(\mathcal X_0)=1\) such that, for every \(x\in\mathcal X_0\) and every
\(\theta\in\Theta\),
\(\Pi_\theta(\cdot| x)\ll \Pi^{\mathrm b}(\cdot| x) \ll F_{T| X}(\cdot| x)\).
Moreover, \(\mathbb E\left[\mathrm{KL}\{\Pi_\theta(\cdot| X)\|\Pi^{\mathrm b}(\cdot| X)\}\right]<\infty\) for every \(\theta\in\Theta\).
\end{assumption}

The first absolute-continuity relation ensures that candidate rules assign mass only to treatments supported by the benchmark, so the benchmark-divergence term is well defined. The second ensures that the benchmark rule, and hence every candidate rule in the class, assigns mass only to treatments supported by the observed data. Equivalently, for \(x\in\mathcal X_0\) and every measurable \(B\subseteq\mathcal T\), \(F_{T|X}(B|x)=0\Rightarrow \Pi^{\mathrm b}(B|x)=0\Rightarrow\Pi_\theta(B|x)=0\).

Under Assumptions \ref{ass:unconfoundedness} and \ref{ass:common-support}, the penalized welfare criterion can be written in terms of observed-data objects as
\begin{align}
    W_{\lambda}(\theta)
    =
    \mathbb E[\mu_\theta(X)]
    -
    \lambda
    \mathbb E\left[
        \mathrm{KL}\{\Pi_\theta(\cdot|X)\|
        \Pi^{\mathrm b}(\cdot|X)\}
    \right],
    \label{eq:criterion-identified}
\end{align}
where
\[
    \mu_\theta(x)
    =
    \int_{\mathcal T} m(t,x)\Pi_\theta(dt|x),
    \qquad
    m(t,x)=\mathbb E[Y|T=t,X=x].
\]
By unconfoundedness, \(m(t,x)=\mathbb E[Y|T=t,X=x]\) equals \(\mathbb E[Y(t)|X=x]\) for \(P_X\)-almost every \(x\) and \(F_{T|X}(\cdot|x)\)-almost every \(t\). The support condition implies \(\Pi_\theta(\cdot|x)\ll F_{T|X}(\cdot|x)\), so the integral defining \(\mu_\theta(x)\) uses only treatment values for which this conditional mean is identified. Thus \(\mu_\theta(x)\) is the mean outcome induced by assigning treatments according to \(\Pi_\theta(\cdot|x)\) for individuals with covariates \(x\).

\subsection{Benchmark-centered assignment classes and regret}\label{subsec:benchmark-centered-class}

The KL penalty suggests a natural benchmark-centered form for randomized assignment rules. To see this, suppose for the moment that, for each covariate value \(x\), the planner could choose any assignment distribution \(\Pi(\cdot|x)\) satisfying \(\Pi(\cdot|x)\ll \Pi^{\mathrm b}(\cdot|x)\). For fixed \(\lambda>0\), the pointwise criterion \(\int_{\mathcal T} m(t,x)\Pi(dt|x)- \lambda \mathrm{KL}\{\Pi(\cdot|x)\|\Pi^{\mathrm b}(\cdot|x)\}\) is strictly concave in $\Pi(\cdot|x)$ and, whenever the denominator is finite, is maximized uniquely by the exponential tilt
\[
    \Pi_{\lambda}^{\mathrm{un}}(dt|x)
    =
    \frac{\exp\{m(t,x)/\lambda\}}
    {\int_{\mathcal T}
        \exp\{m(s,x)/\lambda\}\Pi^{\mathrm b}(ds|x)}
    \Pi^{\mathrm b}(dt|x).
\]
Thus the benchmark rule supplies the baseline assignment distribution, while the exponential factor tilts that distribution toward higher-welfare treatments.

This unrestricted tilt is a useful population reference point, but it is not generally the estimand of the paper. It depends on the unknown function \(m(t,x)\) and may be too flexible or too opaque for implementation. We instead work with finite-dimensional, benchmark-centered classes. A leading example is the exponential-tilt class
\[
    \Pi_\theta(dt| x)
    =
    \frac{\exp\{\theta^\top q(t,x)\}}
    {\int_{\mathcal T}
        \exp\{\theta^\top q(s,x)\}\Pi^{\mathrm b}(ds| x)}
    \Pi^{\mathrm b}(dt| x),
\]
where \(q(t,x)\) is a prespecified vector of assignment features and the normalizing constant is finite for every $\theta\in\Theta$. The class contains the benchmark rule at \(\theta=0\), and nonzero values of \(\theta\) represent covariate-based tilts away from the benchmark. By construction, every rule in the class is absolutely continuous with respect to the benchmark, so the first relation in Assumption~\ref{ass:common-support} holds automatically; the substantive support requirement is the second, that the benchmark assign mass only where the data have support. 

For binary treatment, when both benchmark probabilities are positive, this class reduces to a benchmark-centered logit rule:
\[
    \log
    \frac{\Pi_\theta(\{1\}|x)}
         {\Pi_\theta(\{0\}|x)}
    = \log
    \frac{\Pi^{\mathrm b}(\{1\}|x)}
         {\Pi^{\mathrm b}(\{0\}|x)}
    + \theta^\top\{q(1,x)-q(0,x)\}.
\]
For multivalued treatment, the same construction gives the
benchmark-centered softmax rule used in the application, after normalizing one
treatment arm.

Throughout the paper, the target is the best rule in the specified finite-dimensional class. This best-in-class formulation keeps the assignment rule interpretable and allows the empirical analysis to restrict attention to covariates and treatment features that are relevant for assignment.

\paragraph{Best-in-class rule and regret.}
For fixed \(\lambda\ge 0\), the best assignment rule in the specified class is any solution
\[
    \theta_\lambda^*
    \in
    \arg\max_{\theta\in\Theta} W_\lambda(\theta).
\]
When \(\lambda\) is fixed throughout an argument, we write \(\theta^*\) for \(\theta_\lambda^*\) when doing so causes no confusion. The corresponding assignment rule is \(\Pi_{\theta^*}\). In the efficiency analysis below, we impose conditions under which this best-in-class parameter is unique and interior.

Given an estimator \(\hat\theta\), the main performance object for the
benchmark-improvement problem is penalized-welfare regret,
\[
    R_\lambda(\hat\theta)
    =
    W_\lambda(\theta^*)-W_\lambda(\hat\theta).
\]
This is the loss in the same population criterion from using the estimated rule rather than the oracle best-in-class rule. Since \(W_\lambda(\theta^*)\) is fixed, studying regret is equivalent to studying the centered population value of the chosen rule, with the sign reversed. We use the regret notation because it is the standard welfare-loss measure in the policy-learning literature and because it separates statistical error from the fixed level of the population problem.

The regret is defined relative to the best rule in the specified class
\(\{\Pi_\theta:\theta\in\Theta\}\). The unrestricted maximizer of
\(W_\lambda\) may not belong to this class. The asymptotic theory therefore
characterizes the statistical component of within-class regret, rather than
approximation loss relative to an unrestricted first best.

\begin{remark}[Divergence penalties]
\label{rem:divergence-penalties}
The main efficiency logic is not specific to KL. It extends to smooth strictly convex benchmark-proximity penalties that are finite on the relevant assignment rules and yield an interior, well-curved best-in-class maximizer. We focus on KL because its exponential-tilt characterization motivates the benchmark-centered logit and softmax classes used below, and because it has the information-cost interpretation discussed in the introduction \citep{mattsson2002probabilistic,matejka2015rational}.
\end{remark}

\subsection{The benchmark-proximity weight}
\label{subsec:choosing-lambda}

The benchmark-proximity weight \(\lambda\) is a preference parameter in the planner's objective, not a statistical tuning parameter. Cross-validation or prediction error can help choose nuisance models, but they cannot determine how much institutional value the planner places on remaining close to the benchmark rule. For fixed \(\lambda\ge 0\), the target is the best-in-class rule for that objective. We therefore treat \(\lambda\) as fixed in the theory and report how the estimated rule changes as \(\lambda\) varies.

Equivalently, the objective is the Lagrangian form of a constrained assignment problem. A planner may specify an acceptable average departure from the benchmark and then choose the value of \(\lambda\) that supports the corresponding point on the welfare--divergence frontier. In this interpretation, \(\lambda\) is the shadow value, in outcome units, of relaxing the benchmark proximity constraint. For the unrestricted exponential tilt, average KL divergence from the benchmark is nonincreasing in \(\lambda\), and is strictly decreasing unless the conditional mean outcome is \(\Pi^{\mathrm b}(\cdot|x)\)-almost surely constant for \(P_X\)-almost every \(x\). In a restricted finite-dimensional class, this relationship need not be one-to-one, which is a reason to report the frontier rather than a single
implied value of \(\lambda\).

KL divergence is unit-free because it is an average log ratio of assignment probabilities under the candidate rule and the benchmark rule. It should therefore be read as a measure of relative departure from the benchmark, not as a percentage-point change in assignment probabilities. The multiplier \(\lambda\) converts this unit-free departure into the outcome scale, so \(\lambda\) has the same units as \(Y\). In the simulations and application, we report \(\lambda\) through the dimensionless ratio \(c=\lambda/s_W\), where \(s_W\) is a design-specific scale for conditional welfare contrasts. Smaller values of \(c\) permit more aggressive tilts away from the benchmark. Under the unrestricted tilt, an outcome contrast of size \(s_W\) changes relative log assignment weights by \(1/c\).

\section{Semiparametric Efficiency for Penalized-Welfare Regret}\label{sec:efficiency}

This section develops the statistical theory for the benchmark-improvement problem of Section~\ref{sec:model}: the efficient influence function for the penalized welfare criterion, the semiparametric efficiency bound for the best-in-class assignment parameter, and the second-order limiting distribution of penalized-welfare regret. The organizing distinction is between the fixed-\(\lambda>0\) problem, where the target is a smooth randomized rule that is regular under the conditions below, and the unpenalized boundary, where maximizers are deterministic or sit on the boundary of the class and estimation is nonregular. We fix density notation first, then develop this distinction and the formal results.

For the analysis in this section, \(\lambda>0\) is fixed, and \(\sigma^2(t,x)=\Var(Y|T=t,X=x)\) denotes the conditional variance of the outcome. Let \(\nu\) be a \(\sigma\)-finite measure on \(\mathcal T\) such that \(F_{T|X}(\cdot|x)\ll\nu\) for \(P_X\)-almost every \(x\); in standard cases, \(\nu\) is counting measure for discrete treatments, Lebesgue measure for continuous treatments, and an appropriate mixed dominating measure, such as Lebesgue measure plus counting measure on the relevant atoms, for mixed treatments. By the support chain in Assumption~\ref{ass:common-support}, the benchmark and every candidate rule are then also dominated by \(\nu\). Write \(f(t|x)\), \(\pi_\theta(t|x)\), and \(\pi^{\mathrm b}(t|x)\) for the corresponding densities with respect to \(\nu\). The observed-treatment density \(f(t|x)\) is the generalized propensity score \citep{imbens2000role,imai2004causal}. All integrals over \(\mathcal T\) in this section are with respect to \(\nu\).

The fixed positive benchmark-proximity weight is part of the planner's objective, not an auxiliary statistical smoothing parameter. The planner values proximity to the randomized benchmark because of the manipulation, legitimacy, information, or learning considerations discussed in the introduction. A useful statistical consequence is that the target is a smooth randomized rule rather than a hard assignment rule. Within a smooth finite-dimensional class, and under the interiority, identification, and curvature conditions imposed below, the best-in-class parameter is a regular finite-dimensional target.

The unpenalized boundary is different. When the benchmark-proximity term is omitted, the criterion reduces to ordinary outcome welfare, which is linear in the assignment kernel. Linearity alone does not supply an interior randomized optimum. In the unrestricted problem, an optimizer assigns all probability to treatments with the highest conditional mean. In restricted classes the target remains the within-class maximizer, but classes that allow hard or nearly hard assignments typically place it on the boundary, where the rule is determined by thresholds or argmax comparisons rather than by a smooth benchmark tilt. This is the case studied separately in Section~\ref{sec:regret_upper}.

The source of nonregularity depends on the treatment space. With binary or finite discrete treatments, the value of a fixed deterministic rule is typically a regular treatment-effect functional under overlap. The difficulty comes from estimating the rule. The map from conditional mean outcomes to the best deterministic assignment is nonsmooth: small perturbations of the data-generating law can change which side of a threshold or argmax an individual falls on. This is the source of nonstandard behavior in threshold and hard-classification problems, including cube-root-type asymptotics in related settings \citep{kim1990cube,hirano2009asymptotics,crippa2025regret}.

With continuous treatments, even the value of a fixed deterministic rule is nonregular in the nonparametric model. Such a rule assigns a single treatment level \(g_\theta(x)\) at each covariate value, so its value depends on \(m(t,x)\) only along the graph \(t=g_\theta(x)\). When the observed treatment is continuously distributed, this graph is a lower-dimensional subset of the joint support of \((T,X)\). The resulting point-evaluation problem is not pathwise differentiable without additional structure, and estimating the best deterministic continuous-treatment rule inherits this nonregularity.

These observations explain why the unpenalized deterministic case is not obtained by simply setting \(\lambda=0\) in the positive-weight theory: removing the benchmark-proximity term eliminates the force that keeps the best-in-class rule interior, and the relevant regularity conditions change with the target. Formal pathwise differentiability definitions and deterministic-rule nonregularity results are collected in the online appendix. The main text therefore proceeds in two parts. This section and Section~\ref{sec:welfare_est} study fixed \(\lambda>0\), where regret is locally quadratic. Section~\ref{sec:regret_upper} studies the unpenalized deterministic case through welfare-process bounds rather than a full second-order regret distribution.

The next lemma derives the efficient influence function for the penalized welfare criterion \(W_\lambda(\theta)\) at a fixed assignment rule. Let \(\mathcal P\) denote the nonparametric model of distributions of \(Z\) satisfying Assumptions~\ref{ass:unconfoundedness}--\ref{ass:common-support}
and the moment conditions below. Pathwise differentiability is defined
formally in the online appendix.

\begin{lemma}\label{lm:EIF_formula}
Let Assumptions \ref{ass:unconfoundedness} and \ref{ass:common-support} hold.
Fix $\theta\in\Theta$. Assume that
\begin{enumerate}[label=(\arabic*)]
\item
\(\mathbb E\left[
        \frac{\pi_\theta(T|X)^2}{f(T|X)^2}
        \sigma^2(T,X)
    \right]<\infty\), and 
    \(\mathbb E\left[
        \{\mu_\theta(X)-\lambda
        \mathrm{KL}\{\Pi_\theta(\cdot|X)\|
        \Pi^{\mathrm b}(\cdot|X)\}\}^2
    \right]<\infty\);
\item $m(\cdot,\cdot)$ and $\sigma^2(\cdot,\cdot)$ are measurable and finite.
\end{enumerate}
Then $W_{\lambda}(\theta)$ is pathwise differentiable in $\mathcal P$ with
efficient influence function
\begin{align}
    \varphi_{\lambda,\theta}(Z)
    = \frac{\pi_\theta(T|X)}{f(T|X)}
    \{Y-m(T,X)\} + \mu_\theta(X) - \lambda \mathrm{KL}\{\Pi_\theta(\cdot|X)\|\Pi^{\mathrm b}(\cdot|X)\} - W_{\lambda}(\theta).
    \label{eq:EIF_formula}
\end{align}
Its efficiency bound is
\[
    \Var\{\varphi_{\lambda,\theta}(Z)\}
    = \mathbb E\left[
        \frac{\pi_\theta(T|X)^2}{f(T|X)^2}
        \sigma^2(T,X)
    \right]
    + \Var\{\mu_\theta(X)-\lambda
    \mathrm{KL}\{\Pi_\theta(\cdot|X)\|
    \Pi^{\mathrm b}(\cdot|X)\}\}.
\]
The same expression in \eqref{eq:EIF_formula}, and hence the same efficiency bound, applies in the submodel in which \(f(\cdot|\cdot)\) is known.
\end{lemma}

For the binary treatment case, let \(p(X)=\mathbb P(T=1|X)\), \(m_t(X)=\mathbb E[Y|T=t,X]\), \(\pi_\theta(X)=\Pi_\theta(\{1\}|X)\), and \(\pi^{\mathrm b}(X)=\Pi^{\mathrm b}(\{1\}|X)\). When the benchmark assigns positive probability to both treatment states, the KL term equals
\[
\mathrm{KL}\{\Pi_\theta(\cdot|X)\|\Pi^{\mathrm b}(\cdot|X)\}
= \pi_\theta(X)\log\frac{\pi_\theta(X)}{\pi^{\mathrm b}(X)}
+ \{1-\pi_\theta(X)\}\log\frac{1-\pi_\theta(X)}{1-\pi^{\mathrm b}(X)}.
\]
The efficient influence function in Lemma \ref{lm:EIF_formula} becomes
\begin{align}
    \varphi_{\lambda,\theta}(Z)
    =& \left(\frac{T\{Y-m_1(X)\}}{p(X)} + m_1(X)\right) \pi_\theta(X)+ \left(\frac{(1-T)\{Y-m_0(X)\}}{1-p(X)}
    +m_0(X) \right)\{1-\pi_\theta(X)\}\nonumber\\
    % &+ m_1(X)\pi_\theta(X) + m_0(X)\{1-\pi_\theta(X)\} \nonumber\\
    &- \lambda \left(\pi_\theta(X)\log\frac{\pi_\theta(X)}{\pi^{\mathrm b}(X)} + \{1-\pi_\theta(X)\}\log\frac{1-\pi_\theta(X)}{1-\pi^{\mathrm b}(X)}\right) - W_{\lambda}(\theta).
    \label{eqn:eif_binary}
\end{align}
The first line is the familiar doubly robust score for the outcome welfare component of a randomized assignment rule. The last line accounts for the benchmark-divergence component of the penalized welfare criterion.

We next analyze the efficiency of the best-in-class assignment parameter \(\theta^*\) defined in Section~\ref{sec:model}. The benchmark-centered classes introduced in Section~\ref{subsec:benchmark-centered-class} provide primitive examples in which \(\theta^*\) can be unique and interior after the usual normalizations.

\begin{assumption}\label{ass:penalized_regular}
For the fixed value of \(\lambda>0\), assume: (i) \(\Theta\) is compact; (ii) \(\theta^*\in\operatorname{int}(\Theta)\) is the unique maximizer of \(W_\lambda\) over \(\Theta\); (iii) \(W_\lambda\) is twice continuously differentiable in a neighborhood of \(\theta^*\), and \(H=-\frac{\partial^2 W_\lambda(\theta)}{\partial\theta\partial\theta^\top}\big|_{\theta=\theta^*}\) is positive definite.
\end{assumption}
Appendix~\ref{app:primitive-regularity} shows that the benchmark-centered exponential-tilt classes considered below satisfy Assumption~\ref{ass:penalized_regular} under primitive moment, full-rank, boundary, and curvature conditions on the assignment features. 

The next condition ensures that the influence function in \eqref{eq:EIF_formula} is differentiable in \(\theta\). For any vector \(v\), \(\|v\|\) denotes the Euclidean norm.

\begin{assumption}\label{ass:pol_smooth}
There exists a neighborhood $\mathcal N$ of $\theta^*$ such that, for almost every $(t,x)$, the map $\theta\mapsto\pi_\theta(t|x)$ is differentiable on $\mathcal N$ with derivative \(\frac{\partial\pi_\theta(t|x)}{\partial\theta}\in\mathbb R^p\). Uniformly
in $\theta\in\mathcal N$,
\[ \mathbb E\left[
        \left\|
        \frac{\partial\pi_\theta(T|X)}
             {\partial\theta}
        \right\|^2
        \frac{\sigma^2(T,X)}{f(T|X)^2}
    \right]<\infty, \qquad
    \mathbb E\left[
        \left\|
        \int
        m(t,X)
        \frac{\partial\pi_\theta(t|X)}
             {\partial\theta}
        d\nu(t)
        \right\|^2
    \right]<\infty.
\]
Moreover, for almost every $x$, differentiation and integration are interchangeable for $\mu_\theta(x)$. The map
\(\theta\mapsto \mathrm{KL}\{\Pi_\theta(\cdot|x)\| \Pi^{\mathrm b}(\cdot|x)\}\)
is differentiable on $\mathcal N$ for $P_X$-almost every $x$, and
\[
    \mathbb E\left[
        \sup_{\theta\in\mathcal N}
        \left\|
        \frac{\partial}{\partial\theta}
        \mathrm{KL}\{\Pi_\theta(\cdot|X)\|
        \Pi^{\mathrm b}(\cdot|X)\}
        \right\|^2
    \right]<\infty.
\]
\end{assumption}

The next theorem derives the efficient influence function for the best-in-class assignment parameter and then applies the H\'ajek--Le Cam convolution theorem to characterize the limiting distribution of regular estimators.

\begin{theorem}\label{thm:policy_estimator}
Let Assumptions \ref{ass:unconfoundedness}, \ref{ass:common-support},
\ref{ass:penalized_regular}, and \ref{ass:pol_smooth} hold, together with the
conditions of Lemma \ref{lm:EIF_formula} for every $\theta\in\mathcal N$. The
best-in-class assignment parameter $\theta^*$ is pathwise differentiable with
efficient influence function
\(\mathrm{EIF}_{\theta^*}(Z)
    =
    H^{-1}
    \frac{\partial\varphi_{\lambda,\theta}(Z)}
         {\partial\theta}
    \big|_{\theta=\theta^*}\),
where
\begin{align}
    \frac{\partial\varphi_{\lambda,\theta}(Z)}
         {\partial\theta}
    \bigg|_{\theta=\theta^*}
    =
    &
    \int
    m(t,X)
    \frac{\partial\pi_\theta(t|X)}
         {\partial\theta}
    \bigg|_{\theta=\theta^*}
    d\nu(t)
    +
    \frac{Y-m(T,X)}{f(T|X)}
    \frac{\partial\pi_\theta(T|X)}
         {\partial\theta}
    \bigg|_{\theta=\theta^*}
    \nonumber\\
    &-
    \lambda
    \frac{\partial}{\partial\theta}
    \mathrm{KL}\{\Pi_\theta(\cdot|X)\|
    \Pi^{\mathrm b}(\cdot|X)\}
    \bigg|_{\theta=\theta^*}.
    \label{eq:eif_theta_derivative}
\end{align}
Consequently, the semiparametric efficiency bound for $\theta^*$ is
\[
    V_{\mathrm{eff}}
    =
    H^{-1}
    \Var\left(
        \frac{\partial\varphi_{\lambda,\theta}(Z)}
             {\partial\theta}
        \bigg|_{\theta=\theta^*}
    \right)
    H^{-1}.
\]
If the observed-data treatment density $f(\cdot|\cdot)$ is known, the form
of the efficient influence function remains unchanged. Moreover, for any
regular estimator $\hat\theta$,
\begin{align}
    \sqrt n(\hat\theta-\theta^*)
    \Rightarrow G+U,
    \label{eqn:theta-convolution}
\end{align}
where $G\sim\mathcal N(0,V_{\mathrm{eff}})$ and $U$ is independent of $G$. For regular asymptotically linear estimators, \(U\) is mean zero and Gaussian; write \(\Sigma_U\) for its covariance matrix.
\end{theorem}

In the binary treatment case, the efficient influence function for $\theta^*$
has the following form:
\begin{align*}
    \mathrm{EIF}_{\theta^*}(Z)
    = H^{-1}
    \frac{\partial\pi_\theta(X)}{\partial\theta}
    \bigg|_{\theta=\theta^*}
    \Bigg[
        &
        \frac{T\{Y-m_1(X)\}}{p(X)}
        -
        \frac{(1-T)\{Y-m_0(X)\}}{1-p(X)}
        +
        m_1(X)-m_0(X)
        \\
        &-
        \lambda
        \left\{
        \log\frac{\pi_{\theta^*}(X)}
                 {\pi^{\mathrm b}(X)}
        -
        \log\frac{1-\pi_{\theta^*}(X)}
                 {1-\pi^{\mathrm b}(X)}
        \right\}
    \Bigg].
\end{align*}
The final term is the derivative of the binary KL penalty; it equals the log-odds gap between the best-in-class rule and the benchmark.

\begin{theorem}\label{thm:regret_limit}
Let the assumptions of Theorem \ref{thm:policy_estimator} hold, and suppose the noise component $U$ in \eqref{eqn:theta-convolution} is mean zero with covariance matrix $\Sigma_U$ and finite fourth moments. Then the penalized-welfare regret of $\hat\theta$ satisfies
\[
    nR_\lambda(\hat\theta)\Rightarrow \frac{1}{2}(G+U)^\top H(G+U).
\]
The limiting distribution has mean
\[\frac{1}{2}
    \tr\left[
        H^{1/2}
        \left(V_{\mathrm{eff}}+\Sigma_U\right)
        H^{1/2}
    \right],\]
and variance
\[
    \frac12\sum_{j=1}^p \ell_j^2
    +
    \tr\left(
        H^{1/2}
        V_{\mathrm{eff}}
        H
        \Sigma_U
        H^{1/2}
    \right)
    +
    \frac14
    \Var(U^\top HU),
\]
where $\ell_1\ge\cdots\ge\ell_p\ge0$ are the eigenvalues of $H^{1/2}V_{\mathrm{eff}}H^{1/2}$. If \(U\) is Gaussian, as for regular asymptotically linear estimators, the variance simplifies to \(\frac12\sum_{j=1}^p \tilde \ell_j^2\), where $\tilde \ell_1\ge\cdots\ge\tilde \ell_p\ge0$ are the eigenvalues of $H^{1/2}(V_{\mathrm{eff}}+\Sigma_U)H^{1/2}$.
\end{theorem}

\begin{remark}[Why efficiency matters]
Efficiency matters here because the rule selected from the data has random population value. Although the planner's criterion is \(W_\lambda\), the planner does not implement the oracle rule \(\theta^*\); she implements \(\hat\theta\), and the relevant population value is \(W_\lambda(\hat\theta)\). Regret is just this random value centered at the oracle best-in-class value and written as a loss. The local geometry makes the role of efficiency transparent. Around the interior optimum,
\[
    W_\lambda(\hat\theta)
    =
    W_\lambda(\theta^*)
    -
    \frac12(\hat\theta-\theta^*)'H(\hat\theta-\theta^*)
    +
    o_p(\|\hat\theta-\theta^*\|^2).
\]
Thus two estimators with the same target differ in welfare through the dispersion of their rule-estimation error. An inefficient estimator carries extra first-order noise, so the population value of the selected rule is more variable and, because the loss is quadratic, lower on average at the \(1/n\) scale. In this sense efficiency is not only an inference concept: it means that the data-driven rule is more tightly concentrated near the best-in-class rule and has smaller expected welfare loss.
\end{remark}

Theorem~\ref{thm:regret_limit} therefore turns the usual semiparametric efficiency comparison into a welfare comparison. If two regular estimators have first-order covariance matrices \(V_1\preceq V_2\), then the estimator with covariance \(V_1\) has weakly smaller mean limiting regret. The next section applies this implication to true-propensity IPW, estimated-propensity IPW, and doubly robust estimation.

\section{Penalized-Welfare Estimators and Regret} \label{sec:welfare_est}

The previous section characterized the semiparametric efficiency bound for the best-in-class assignment parameter and the associated regret distribution. This section studies three estimators of the penalized welfare criterion: IPW using the known observed-treatment density, IPW using an estimated observed-treatment density, and doubly robust estimation. All three estimators maximize an estimate of the same criterion \(W_\lambda(\theta)\). They differ only in how the outcome-welfare component is estimated; the benchmark-divergence component is common across methods because \(\Pi^{\mathrm b}\) is known. The comparison therefore isolates how the estimation of counterfactual outcome welfare affects assignment-rule efficiency and regret.

\subsection{IPW estimator with true propensity}
\label{subsec:ipw_tp}

By Assumptions~\ref{ass:unconfoundedness}--\ref{ass:common-support} and
iterated expectations, the outcome-welfare component of \(W_\lambda(\theta)\)
admits the IPW representation
\[
    \mathbb E\left[
        \int_{\mathcal T} Y(t)\Pi_\theta(dt|X)
    \right]
    =
    \mathbb E\left[
        \frac{\pi_\theta(T|X)Y}{f(T|X)}
    \right].
\]
Therefore, when the observed-data propensity \(f(\cdot|\cdot)\) is known, the
penalized welfare criterion can be estimated by
\begin{align*}
    \widehat W_\lambda^{tp}(\theta)
    =
    \frac{1}{n}\sum_{i=1}^n
    \left[
        \frac{\pi_\theta(T_i|X_i)Y_i}{f(T_i|X_i)}
        -
        \lambda\,\mathrm{KL}\{\Pi_\theta(\cdot|X_i)\|
        \Pi^{\mathrm b}(\cdot|X_i)\}
    \right],
\end{align*}
and the corresponding assignment-rule estimator is
\(\hat\theta^{tp}
    =
    \arg\max_{\theta\in\Theta}
    \widehat W_\lambda^{tp}(\theta)\). Here, the superscript ``\textit{tp}'' stands for true propensity.

\begin{assumption}\label{ass:ipw_true}
There exists \(M<\infty\) such that \(\sup_{t\in\mathcal T}|Y(t)|\le M\)
almost surely. There exists \(\underline f>0\) such that, for
\(P_X\)-almost every \(x\), \(f(t|x)\ge \underline f \) for \(\nu\)-almost every \(t\) satisfying \(\pi^{\mathrm b}(t|x)>0\).
\end{assumption}

Assumption~\ref{ass:ipw_true} imposes bounded potential outcomes and overlap on
the support relevant for candidate rules. Since
\(\Pi_\theta(\cdot|x)\ll\Pi^{\mathrm b}(\cdot|x)\), the IPW ratio is used only
where the benchmark assigns positive density.

\begin{theorem}
\label{thm:ipw_tp}
Under Assumptions~\ref{ass:unconfoundedness},
\ref{ass:common-support}, \ref{ass:penalized_regular},
\ref{ass:pol_smooth}, and~\ref{ass:ipw_true}, together with the conditions in
Lemma~\ref{lm:EIF_formula} for \(\theta\) in a neighborhood of \(\theta^*\),
the estimator \(\hat\theta^{tp}\) satisfies
\[
    \sqrt n(\hat\theta^{tp}-\theta^*)
    \Rightarrow
    G+U^{tp},
    \qquad
    nR_\lambda(\hat\theta^{tp})
    \Rightarrow
    \frac12
    (G+U^{tp})^\top H(G+U^{tp}),
\]
where \(G\sim \mathcal N(0,V_{\mathrm{eff}})\) and
\(U^{tp}\sim \mathcal N(0,\Sigma_U^{tp})\) is independent of \(G\), with
\[
    \Sigma_U^{tp}
    =
    H^{-1}
    \Var\left(
        \frac{m(T,X)}{f(T|X)}
        \frac{\partial\pi_\theta(T|X)}{\partial\theta}
        \bigg|_{\theta=\theta^*}
        -
        \frac{\partial\mu_\theta(X)}{\partial\theta}
        \bigg|_{\theta=\theta^*}
    \right)
    H^{-1}.
\]
The limiting regret distribution has mean
\(\frac{1}{2}\tr\left[H^{1/2}(V_{\mathrm{eff}}+\Sigma_U^{tp})H^{1/2}\right]\)
and variance \(\frac{1}{2}\sum_{j=1}^p(\tilde\ell_j^{tp})^2\), where
\(\tilde\ell_1^{tp}\ge\cdots\ge\tilde\ell_p^{tp}\ge0\) are the eigenvalues of
\(H^{1/2}(V_{\mathrm{eff}}+\Sigma_U^{tp})H^{1/2}\).
\end{theorem}

The matrix \(\Sigma_U^{tp}\) is positive semidefinite. It is zero if and only if \(\frac{m(T,X)}{f(T|X)}\frac{\partial\pi_\theta(T|X)}{\partial\theta}\big|_{\theta=\theta^*}\) is almost surely equal, conditional on \(X\), to its conditional mean \(\frac{\partial\mu_\theta(X)}{\partial\theta}\big|_{\theta=\theta^*}\). Except in this degenerate case, IPW using the true propensity is inefficient. The extra term is treatment-assignment variation in the IPW score: it has conditional mean zero given \(X\), is orthogonal to the efficient component, and therefore adds variance. The KL penalty contributes no additional noise. Its score contribution is identical across the criteria and cancels from the difference of scores, so \(\Sigma_U^{tp}\) involves only the outcome-welfare component. The benchmark-proximity weight \(\lambda\) still affects \(\theta^*\), the efficient component, and \(H\), but the extra inefficiency of true-propensity IPW comes entirely from the outcome-welfare part.

The inefficiency of true-propensity IPW is the HIR phenomenon in this assignment-rule setting \citep{hirano2003efficient,chen2008semiparametric,ai2021unified}. Here the implication is decision-theoretic: the extra variance does not merely widen the distribution of the assignment-parameter estimator; it raises the mean of limiting regret.

\subsection{IPW estimator with estimated propensity}
\label{subsec:ipw_ep}

Let \(\omega(t,x)=1/f(t|x)\) denote the inverse observed-treatment density, so
that
\[
    \mathbb E\left[
        \int_{\mathcal T}Y(t)\Pi_\theta(dt|X)
    \right]
    =
    \mathbb E[\pi_\theta(T|X)\omega(T,X)Y].
\]
This motivates a two-step estimator that first estimates \(\omega\) and then
plugs the estimated weights into the penalized welfare criterion.

The estimation of \(\omega\) builds on the balancing approach of
\citet{ai2021unified}, whose weighting function is \(f_T(t)/f(t|x)\), with
\(f_T\) denoting the marginal treatment density. In our criterion, the rule
density \(\pi_\theta(T|X)\) enters as a known multiplier of the weight, so the
relevant weight is the raw inverse density \(1/f(T|X)\). This weight is
characterized by the balancing condition that, for all integrable functions
\(u(T)\) and \(v(X)\),
\[
    \mathbb E[\omega(T,X)u(T)v(X)]
    =
    \left(\int_{\mathcal T} u(t)d\nu(t)\right)\mathbb E[v(X)].
\]
Thus the balancing condition integrates \(u\) against the dominating measure
\(\nu\), rather than against the marginal treatment distribution.

We approximate the function spaces for \(u\) and \(v\) by finite-dimensional
sieves
\(u_{K_1}(T)=(u_{K_1,1}(T),\ldots,u_{K_1,K_1}(T))^\top\) and
\(v_{K_2}(X)=(v_{K_2,1}(X),\ldots,v_{K_2,K_2}(X))^\top\), with
\(K=K_1K_2\). When \(T\) has finite discrete support, only the covariate basis
needs to grow. The entropy-tilting program is
\begin{align*}
    \max_{\{\omega_i>0\}_{i=1}^n}
    \quad
    &-\sum_{i=1}^n \omega_i\log\omega_i
    \\
    \text{s.t.}\quad
    &
    \frac1n\sum_{i=1}^n
    \omega_i u_{K_1}(T_i)v_{K_2}(X_i)^\top
    =
    \left(\int_{\mathcal T} u_{K_1}(t)d\nu(t)\right)
    \left(\frac1n\sum_{i=1}^n v_{K_2}(X_i)\right)^\top .
\end{align*}
Let \(\rho(v)=-\exp(-v-1)\), so that \(\rho'(v)=\exp(-v-1)>0\). The estimated
weight is
\begin{align*}
    \hat\omega_K(T_i,X_i)
    &=
    \rho'\{u_{K_1}(T_i)^\top
    \hat\Lambda_{K_1\times K_2}v_{K_2}(X_i)\},
    \\
    \hat\Lambda_{K_1\times K_2}
    &=
    \arg\max_{\Lambda}
    \frac1n\sum_{i=1}^n
    \rho\{u_{K_1}(T_i)^\top\Lambda v_{K_2}(X_i)\}
    -
    \left(\int_{\mathcal T} u_{K_1}(t)d\nu(t)\right)^\top
    \Lambda
    \left(\frac1n\sum_{i=1}^n v_{K_2}(X_i)\right).
\end{align*}

\begin{assumption}
\label{ass:est_weight}
\begin{enumerate}[label=(\arabic*)]
\item The supports \(\mathcal X\) and \(\mathcal T\) are compact,
\(\nu(\mathcal T)<\infty\), and there exist \(0<\underline f<\overline f<\infty\) such that \(\underline f\le f(t|x)\le \overline f\) for \(\nu\otimes P_X\)-almost every \((t,x)\).

\item There exist \(\Lambda_{K_1\times K_2}\in\mathbb R^{K_1\times K_2}\)
and \(\alpha>0\) such that
\[
    \left\|
    (\rho')^{-1}\{\omega(t,x)\}
    -
    u_{K_1}(t)^\top
    \Lambda_{K_1\times K_2}
    v_{K_2}(x)
    \right\|_\infty
    =
    O(K^{-\alpha}).
\]

\item The bases include constants. The smallest eigenvalues of \(\int_{\mathcal T} u_{K_1}(t)u_{K_1}(t)^\top d\nu(t)\) and \(\mathbb E[v_{K_2}(X)v_{K_2}(X)^\top]\) are bounded away from zero uniformly in \(K_1\) and \(K_2\).

\item There exist sequences \(\zeta_1(K_1)\) and \(\zeta_2(K_2)\) with \(\sup_{t\in\mathcal T}\|u_{K_1}(t)\|\le \zeta_1(K_1)\), and
    \(\sup_{x\in\mathcal X}\|v_{K_2}(x)\|\le \zeta_2(K_2)\).
Let \(\zeta(K)=\zeta_1(K_1)\zeta_2(K_2)\). The sieve dimensions satisfy
\(\zeta(K)K/\sqrt n\to 0\), and \(\sqrt n K^{-\alpha}\to0\).

\item The same tensor-product sieve approximates the score weights in the
following sense:
\[
    \sqrt K
    \max_{1\le j\le p}
    \inf_{\Lambda\in\mathbb R^{K_1\times K_2}}
    \left\|
        m(t,x)
    \frac{\partial\pi_\theta(t|x)}{\partial\theta_j}
        \right|_{\theta=\theta^*}
        -
        u_{K_1}(t)^\top\Lambda v_{K_2}(x)
    \bigg\|_{L_2(\nu\otimes P_X)}
    \to 0.
\]
\end{enumerate}
\end{assumption}

Assumption~\ref{ass:est_weight} adapts the entropy-balancing conditions of \citet{ai2021unified} to the present general-treatment assignment setting. Condition~(1) strengthens the overlap requirement in Assumption~\ref{ass:ipw_true} to the full support used by the entropy-balancing construction. Conditions~(2)--(4) give the estimated-weight rate in Lemma~\ref{lm:weight_rate}. Condition~(5) is used only for the local score expansion in Theorem~\ref{thm:ipw_ep}: it requires the same tensor-product sieve to approximate the components of \(m(t,x)\partial\pi_\theta(t|x)/\partial\theta\) at \(\theta=\theta^*\) fast enough for the balancing equations to remove the treatment-assignment score component. For standard polynomial, spline, or wavelet bases, this is a smoothness requirement on that product.

\begin{lemma}
\label{lm:weight_rate}
Under Assumption~\ref{ass:ipw_true} and parts (1)--(4) of Assumption~\ref{ass:est_weight},
\[
    \int_{\mathcal T\times\mathcal X}
    |\hat\omega_K(t,x)-\omega(t,x)|^2
    dF_{T,X}(t,x)
    = O_p(K/n),
    \qquad
    \frac1n\sum_{i=1}^n
    |\hat\omega_K(T_i,X_i)-\omega(T_i,X_i)|^2 = O_p(K/n).
\]
\end{lemma}

The proof of Theorem~\ref{thm:ipw_ep} combines this rate with the exact balancing equations, as shown in the appendix, to obtain the local score expansion for the estimated-propensity criterion. The estimated-propensity (``\textit{ep}'') criterion is
\begin{align*}
    \widehat W_\lambda^{ep}(\theta)
    =
    \frac1n\sum_{i=1}^n
    \left[
        \hat\omega_K(T_i,X_i)\pi_\theta(T_i|X_i)Y_i
        -
        \lambda\,
        \mathrm{KL}\{\Pi_\theta(\cdot|X_i)\|
        \Pi^{\mathrm b}(\cdot|X_i)\}
    \right],
\end{align*}
with \(\hat\theta^{ep}=\arg\max_{\theta\in\Theta}\widehat W_\lambda^{ep}(\theta)\).

\begin{theorem}
\label{thm:ipw_ep}
Under Assumptions~\ref{ass:unconfoundedness}, \ref{ass:common-support},
\ref{ass:penalized_regular}, \ref{ass:pol_smooth}, \ref{ass:ipw_true},
and \ref{ass:est_weight}, together with the conditions in
Lemma~\ref{lm:EIF_formula} for \(\theta\) in a neighborhood of
\(\theta^*\), the estimator \(\hat\theta^{ep}\) satisfies
\[
    \sqrt n(\hat\theta^{ep}-\theta^*)
    \Rightarrow
    G,
    \qquad
    nR_\lambda(\hat\theta^{ep})
    \Rightarrow
    \frac12G^\top HG,
\]
where \(G\sim\mathcal N(0,V_{\mathrm{eff}})\). Therefore, the IPW estimator with
estimated propensity attains efficient penalized-welfare regret.
\end{theorem}

Theorem~\ref{thm:ipw_ep} is the assignment-rule analogue of the HIR phenomenon:
even when the observed-treatment density is known, replacing it with suitable
estimated weights removes the extra variance component in
Theorem~\ref{thm:ipw_tp}. The result should not be read as a recommendation to
estimate the known propensity by an arbitrary low-dimensional model. Efficiency
requires balancing restrictions rich enough to absorb the orthogonal
treatment-assignment component in the true-propensity score asymptotically.

\subsection{Doubly robust estimator}
\label{subsec:dr}

The efficient influence function in Lemma~\ref{lm:EIF_formula} yields the
doubly robust representation
\begin{align*}
    W_\lambda(\theta)
    =
    \mathbb E\left[
        \omega(T,X)\pi_\theta(T|X)\{Y-m(T,X)\}
        +
        \mu_\theta(X)
        -
        \lambda
        \mathrm{KL}\{\Pi_\theta(\cdot|X)\|
        \Pi^{\mathrm b}(\cdot|X)\}
    \right],
\end{align*}
where \(\omega(t,x)=1/f(t|x)\). The outcome-welfare component is the usual
doubly robust expression: it remains correctly centered if either the inverse
density or the outcome regression is correctly specified. The
benchmark-divergence term enters directly because it depends only on \(X\), the
candidate rule, and the known benchmark rule.

We use cross-fitting. Partition the sample into \(L\ge2\) folds
\(\{\mathcal I_\ell\}_{\ell=1}^L\). For each fold \(\ell\), estimate the
nuisance functions on the complement \(\mathcal I_\ell^c\), obtaining
\(\hat m^{(-\ell)}\) and \(\hat\omega^{(-\ell)}\), and evaluate them on
\(\mathcal I_\ell\). For \(x\in\mathcal X\), define
\[
    \hat\mu_\theta^{(-\ell)}(x)
    =
    \int_{\mathcal T}
    \hat m^{(-\ell)}(t,x)\pi_\theta(t|x)d\nu(t).
\]
The doubly robust penalized-welfare criterion is
\begin{align*}
    \widehat W_\lambda^{dr}(\theta)
    =
    \frac1n
    \sum_{\ell=1}^L
    \sum_{i\in\mathcal I_\ell}
    \Big[
        &
        \hat\omega^{(-\ell)}(T_i,X_i)\pi_\theta(T_i|X_i)
        \{Y_i-\hat m^{(-\ell)}(T_i,X_i)\}
        +
        \hat\mu_\theta^{(-\ell)}(X_i)
        \\
        &-
        \lambda\,
        \mathrm{KL}\{\Pi_\theta(\cdot|X_i)\|
        \Pi^{\mathrm b}(\cdot|X_i)\}
    \Big],
\end{align*}
and \(\hat\theta^{dr}=\arg\max_{\theta\in\Theta}\widehat W_\lambda^{dr}(\theta)\). Here, the superscript ``\textit{dr}'' stands for doubly robust. Let \(\|\cdot\|_{L_2}\) denote the \(L_2\)-norm under the distribution of \((T,X)\). We impose the following conditions.

\begin{assumption}\label{ass:dr}
For each fold \(\ell\), the nuisance estimators trained on
\(\mathcal I_\ell^c\) satisfy the following conditions, uniformly over
\(\ell=1,\ldots,L\). 
\begin{enumerate}[label=(\arabic*)]
\item There exist \(\rho_\omega,\rho_m>0\), with
\(\rho_\omega+\rho_m\ge 1/2\), such that
\(\|\hat\omega^{(-\ell)}-\omega\|_{L_2}=o_p(n^{-\rho_\omega})\), and \(\|\hat m^{(-\ell)}-m\|_{L_2}=o_p(n^{-\rho_m})\).
\item With probability approaching one, \(\hat\omega^{(-\ell)}\) and
\(\hat m^{(-\ell)}\) are uniformly bounded over \((t,x)\).
\end{enumerate}
\end{assumption}

Assumption~\ref{ass:dr}(1) is the usual product-rate condition for Neyman
orthogonality, and Assumption~\ref{ass:dr}(2) rules out unstable first-stage
extrapolation. The conditional mean \(m\) may be estimated by sieve methods
\citep{chen2007large}, local polynomial methods \citep{calonico2018effect},
partitioning methods \citep{cattaneo2024uniform}, or machine learning methods
\citep{chernozhukov2018double}. The inverse density \(\omega\) may be estimated
by the balancing method above or by first estimating the conditional density
\(f\) and then taking its inverse \citep{cattaneo2024boundary,
colangelo2025double}.

\begin{theorem}
\label{thm:dr}
Under Assumptions~\ref{ass:unconfoundedness}, \ref{ass:common-support},
\ref{ass:penalized_regular}, \ref{ass:pol_smooth}, \ref{ass:ipw_true}, and
\ref{ass:dr}, together with the conditions in Lemma~\ref{lm:EIF_formula} for
\(\theta\) in a neighborhood of \(\theta^*\), the estimator
\(\hat\theta^{dr}\) satisfies
\[
    \sqrt n(\hat\theta^{dr}-\theta^*)
    \Rightarrow
    G,
    \qquad
    nR_\lambda(\hat\theta^{dr})
    \Rightarrow
    \frac12G^\top HG,
\]
where \(G\sim \mathcal N(0,V_{\mathrm{eff}})\). Therefore, the doubly robust
estimator attains efficient penalized-welfare regret.
\end{theorem}

Neyman orthogonality eliminates first-order sensitivity to the nuisance
estimators, so first-stage errors enter only through a second-order remainder.
Cross-fitting, the product-rate condition in Assumption~\ref{ass:dr}, and the
smoothness of the assignment class deliver the required local expansion of
\(\widehat W_\lambda^{dr}\). Consequently, the doubly robust estimator has the
efficient influence function derived in Theorem~\ref{thm:policy_estimator} and
achieves the efficient penalized-welfare regret distribution.

Together, Theorems~\ref{thm:ipw_tp}--\ref{thm:dr} establish the HIR phenomenon
for penalized-welfare assignment-rule estimation. True-propensity IPW generally
carries an additional positive-semidefinite variance component. Estimated-propensity IPW and doubly robust estimation remove that component, attain the
semiparametric efficiency bound for \(\theta^*\), and therefore attain the
efficient regret distribution.

\section{Welfare-Process Bounds in the Unpenalized Case}
\label{sec:regret_upper}

Sections~\ref{sec:efficiency} and~\ref{sec:welfare_est} study the
benchmark-improvement problem with fixed \(\lambda>0\). This section asks
whether the same efficiency force appears in the standard unpenalized
deterministic setting. We focus on binary treatment and a class of deterministic
rules with finite VC dimension. Unlike the positive-penalty case, we do not
derive a full second-order distribution for regret. Instead, we compare the
Gaussian welfare processes that enter the usual basic-inequality upper bound
for regret. The message is that the HIR phenomenon is not an artifact of the
smooth penalized criterion: true-propensity IPW still carries an orthogonal
assignment-noise component that estimated-propensity IPW removes.

Represent an assignment rule by a function
\(\pi:\mathcal X\to\{0,1\}\), where \(\pi(x)=1\) assigns treatment and
\(\pi(x)=0\) assigns control. Let \(\mathbf{\Pi}\) be a class of such rules, not indexed by the finite-dimensional parameter \(\theta\).\footnote{The
analysis extends to randomized rules \(\pi:\mathcal X\to[0,1]\), with VC
subgraph dimension replacing VC dimension.} In this section only,
\(W(\pi)\) denotes unpenalized outcome welfare:
\[
    W(\pi)
    =
    \mathbb E\left[
        Y(1)\pi(X)+Y(0)\{1-\pi(X)\}
    \right]
    =
    \mathbb E\left[
        m_1(X)\pi(X)+m_0(X)\{1-\pi(X)\}
    \right].
\]

For an estimator \(\widehat W(\pi)\) of \(W(\pi)\), let
\[
    \hat\pi\in\arg\max_{\pi\in\mathbf{\Pi}}\widehat W(\pi),
    \qquad
    \pi^*\in\arg\max_{\pi\in\mathbf{\Pi}}W(\pi),
\]
and define regret by \(R(\hat\pi)=W(\pi^*)-W(\hat\pi)\).
The usual basic inequality gives
\begin{align}
    R(\hat\pi)
    &\le
    \big(W(\pi^*)-\widehat W(\pi^*)\big)
    +
    \underbrace{
    \big(\widehat W(\pi^*)-\widehat W(\hat\pi)\big)
    }_{\le 0}
    +
    \big(\widehat W(\hat\pi)-W(\hat\pi)\big)
    \nonumber\\
    &\le
    2\sup_{\pi\in\mathbf{\Pi}}
    |\widehat W(\pi)-W(\pi)|.
    \label{eqn:basic-inequality}
\end{align}
The object studied in this section is therefore the welfare-process bound on
the right-hand side of \eqref{eqn:basic-inequality}, not regret itself. If
\(\sqrt n\{\widehat W(\pi)-W(\pi)\}\) converges weakly in
\(\ell^\infty(\mathbf{\Pi})\) to a centered tight Gaussian process \(G(\pi)\),
then
\[
    \sup_{\pi\in\mathbf{\Pi}}
    \left|
        \sqrt n\{\widehat W(\pi)-W(\pi)\}
    \right|
    \Rightarrow
    \sup_{\pi\in\mathbf{\Pi}}|G(\pi)|.
\]
The mean of this limiting supremum is informative about the asymptotic size of
the regret upper bound, although it is not the mean of regret.

We compare two IPW estimators of \(W(\pi)\). The first uses the true propensity score \(p(X)\), with \(\omega_1(X)=1/{p(X)}\), and \(\omega_0(X)=1/{(1-p(X))}\). The second replaces these weights by estimated weights
\(\hat\omega_1(X)\) and \(\hat\omega_0(X)\), obtained from the binary
specialization of the estimated-propensity method in
Section~\ref{subsec:ipw_ep}:
\begin{align*}
    \widehat W^{tp}(\pi)
    &=
    \frac1n\sum_{i=1}^n
    \left[
        \omega_1(X_i)T_iY_i\pi(X_i)
        +
        \omega_0(X_i)(1-T_i)Y_i\{1-\pi(X_i)\}
    \right],
    \\
    \widehat W^{ep}(\pi)
    &=
    \frac1n\sum_{i=1}^n
    \left[
        \hat\omega_1(X_i)T_iY_i\pi(X_i)
        +
        \hat\omega_0(X_i)(1-T_i)Y_i\{1-\pi(X_i)\}
    \right].
\end{align*}

\begin{assumption}\label{ass:policyclass}
(i) There exists \(\kappa>0\) such that
\(\kappa\le p(X)=\mathbb P(T=1|X)\le 1-\kappa\) almost surely, and there exists \(M<\infty\) such that \(|Y(1)|,|Y(0)|\le M\) almost surely; (ii) the class of assignment rules
\(\mathbf{\Pi}\) has finite VC dimension \(\mathrm{VC}(\mathbf{\Pi})<\infty\).
\end{assumption}
\begin{assumption}\label{ass:det-balance}
Let \(V_K=\mathrm{span}\{v_{K_2,1},\ldots,v_{K_2,K_2}\}\) be the
covariate sieve used by the binary estimated weights. The sieve approximates
the assignment-weighted outcome regressions uniformly over
\(\mathbf\Pi\):
\[
    \sqrt{K\log n}\,
    \sup_{\pi\in\mathbf\Pi}
    \max\left\{
    \inf_{v\in V_K}
    \|m_1\pi-v\|_{L_2(P_X)},
    \inf_{v\in V_K}
    \|m_0(1-\pi)-v\|_{L_2(P_X)}
    \right\}
    \to0,
\]
and
\(\frac{\zeta(K)K\log n}{\sqrt n}\to0 \).
\end{assumption}

\begin{theorem}\label{thm:ipw_general}
Under Assumptions~\ref{ass:unconfoundedness} and~\ref{ass:policyclass},
\[
    \sup_{\pi\in\mathbf{\Pi}}
    \left|
        \sqrt n\{\widehat W^{tp}(\pi)-W(\pi)\}
    \right|
    =
    \sup_{\pi\in\mathbf{\Pi}}
    \left|
        \frac1{\sqrt n}
        \sum_{i=1}^n
        \varphi^{tp}_\pi(Z_i)
    \right|
    \Rightarrow
    \sup_{\pi\in\mathbf{\Pi}}|G_{tp}(\pi)|.
\]
If, in addition, the estimated weights are constructed under
Assumption~\ref{ass:est_weight}(1)--(4) and
Assumption~\ref{ass:det-balance} holds, then
\[
    \sup_{\pi\in\mathbf{\Pi}}
    \left|
        \sqrt n\{\widehat W^{ep}(\pi)-W(\pi)\}
    \right|
    =
    \sup_{\pi\in\mathbf{\Pi}}
    \left|
        \frac1{\sqrt n}
        \sum_{i=1}^n
        \varphi^{ep}_\pi(Z_i)
    \right|
    +
    o_p(1)
    \Rightarrow
    \sup_{\pi\in\mathbf{\Pi}}|G_{ep}(\pi)|.
\]
Here \(G_{tp}\) and \(G_{ep}\) are centered tight Gaussian processes on
\(\mathbf{\Pi}\) with covariance functions induced by
\begin{align}
    \varphi^{tp}_\pi(Z)
    &=
    \left[
        \omega_1(X)TY
        -
        \omega_0(X)(1-T)Y
    \right]\pi(X)
    +
    \omega_0(X)(1-T)Y
    -
    W(\pi),
    \nonumber\\
    \varphi^{ep}_\pi(Z)
    &=
    \left[
        \omega_1(X)T\{Y-m_1(X)\}
        -
        \omega_0(X)(1-T)\{Y-m_0(X)\}
        +
        m_1(X)-m_0(X)
    \right]\pi(X)
    \nonumber\\
    &\quad
    +
    \omega_0(X)(1-T)\{Y-m_0(X)\}
    +
    m_0(X)
    -
    W(\pi).
    \label{eqn:gaussian}
\end{align}
Moreover, the expected limiting welfare-process bound is weakly smaller under estimated-propensity weighting:
\[
    \mathbb E\left[
        \sup_{\pi\in\mathbf{\Pi}}|G_{ep}(\pi)|
    \right]
    \le
    \mathbb E\left[
        \sup_{\pi\in\mathbf{\Pi}}|G_{tp}(\pi)|
    \right].
\]
\end{theorem}
Assumption~\ref{ass:det-balance} is used only for the
estimated-propensity conclusion. It requires the covariate-balancing sieve to
approximate the assignment-weighted outcome regressions uniformly over
\(\mathbf\Pi\); for deterministic rules this is a joint restriction on the rule
class and the sieve, not merely a smoothness condition.

Theorem~\ref{thm:ipw_general} shows that the relationship between the two
welfare processes is exact, not merely a covariance comparison. The
estimated-propensity process is driven by the efficient welfare influence
function for fixed binary assignment rules. The true-propensity process adds
the treatment-assignment component
\begin{align*}
    \psi_\pi(Z)=&\varphi^{tp}_\pi(Z)-\varphi^{ep}_\pi(Z)\\
 =&\big[
        m_1(X)\{\omega_1(X)T-1\}
        -
        m_0(X)\{\omega_0(X)(1-T)-1\}
    \big]\pi(X)
    +
    m_0(X)\{\omega_0(X)(1-T)-1\}.
\end{align*}
Because \(\mathbb E[\omega_1(X)T|X]=\mathbb E[\omega_0(X)(1-T)|X]=1\), the process \(\varphi^{tp}_\pi-\varphi^{ep}_\pi\) has conditional mean zero given \(X\). It is also
uncorrelated with \(\varphi^{ep}_{\pi'}\) for every
\(\pi,\pi'\in\mathbf{\Pi}\): the residual part of \(\varphi^{ep}_{\pi'}\) has mean zero given \((T,X)\), while the remaining part is \(X\)-measurable. Hence \(G_{tp}\overset{d}{=}G_{ep}+U^{tp}\), where \(U^{tp}\) is a centered Gaussian process, independent of \(G_{ep}\), with covariance function \(\mathbb E[\psi_\pi(Z)\psi_{\pi'}(Z)]\).
This is the process analogue of the orthogonal noise component in
Theorem~\ref{thm:ipw_tp}.

The comparison of expected suprema follows because adding an independent
mean-zero Gaussian noise process cannot decrease the expectation of a convex functional of the process. Conditional Jensen's inequality gives the final display in Theorem~\ref{thm:ipw_general}; measurability details are handled in the appendix by a pointwise measurability convention for $\mathbf\Pi$ and by taking separable modifications of the limiting Gaussian processes.

\section{Empirically Calibrated Simulations}\label{sec:simulation}

This section uses calibrated simulations to illustrate the regret theory of
Section~\ref{sec:welfare_est}. Theorems~\ref{thm:ipw_tp}--\ref{thm:dr}
predict that normalized regret, \(nR_\lambda(\hat\theta)\), has a quadratic
limit. True-propensity IPW carries an additional orthogonal component, while
estimated-propensity IPW and doubly robust estimation attain the efficient
limit. The simulations provide a controlled comparison of this prediction:
true-propensity IPW should have larger mean normalized regret, while the two
efficient criteria should behave similarly as \(n\) grows.

The design is calibrated to the Job Training Partnership Act (JTPA) data,
widely used in the statistical treatment-choice literature
\citep{kitagawa2018should,kitagawa2021equality,mbakop2021model,crippa2025regret}.
Let \((Y^*,T^*,X^*)\) denote the original data and \((Y,T,X)\) the simulated
variables. The outcome \(Y^*\) is post-treatment earnings in thousands of
dollars, and \(T^*\) is the treatment indicator. We remove columns with
missing values, expand site indicators, standardize covariates, and use two
assignment covariates: education and pre-program earnings, denoted
\(\texttt{edu}\) and \(\texttt{prevearn}\). To avoid extreme covariate values,
we trim the pool to the 5th--95th percentiles of these two variables and draw
samples of size \(n\in\{500,1000,1500\}\) with replacement.

Using the 8{,}192-observation JTPA analysis file, we fit arm-specific random
forests and treat the fitted functions \(m_0^*(\cdot)\) and \(m_1^*(\cdot)\)
as the true conditional means. Potential outcomes are generated as
\(Y(t)=m_t^*(X)+\varepsilon_t\), with symmetric uniform shocks whose variances
match the arm-specific residual variances. This preserves nonlinear
heterogeneity while giving oracle access to welfare and regret. Because the
DGP uses random-forest fits as the truth, it may be relatively favorable to
the DR criterion with random-forest nuisances; the IPW comparison between true
and estimated propensities does not use outcome regression and is not driven by
this feature.

The observed-data propensity and the benchmark rule are distinct. Treatment is
assigned with
\[
    p(x)=\frac{\exp\{0.5-0.5\,\texttt{edu}\}}
    {1+\exp\{0.5-0.5\,\texttt{edu}\}},
\]
clipped to \([0.05,0.95]\); the average simulated assignment probability is
0.616. The benchmark in the planner's criterion is the uniform lottery,
\(\pi^{\mathrm b}(1\mid x)=0.5\). The benchmark-centered tilt then reduces to
\[
    \pi_\theta(1\mid x)
    =
    \frac{\exp\{\theta_0+\theta_1\texttt{edu}
    +\theta_2\texttt{prevearn}\}}
    {1+\exp\{\theta_0+\theta_1\texttt{edu}
    +\theta_2\texttt{prevearn}\}},
\]
so \(\theta=0\) reproduces the benchmark lottery.

\begin{figure}[p]
    \centering
    \includegraphics[width=\linewidth]{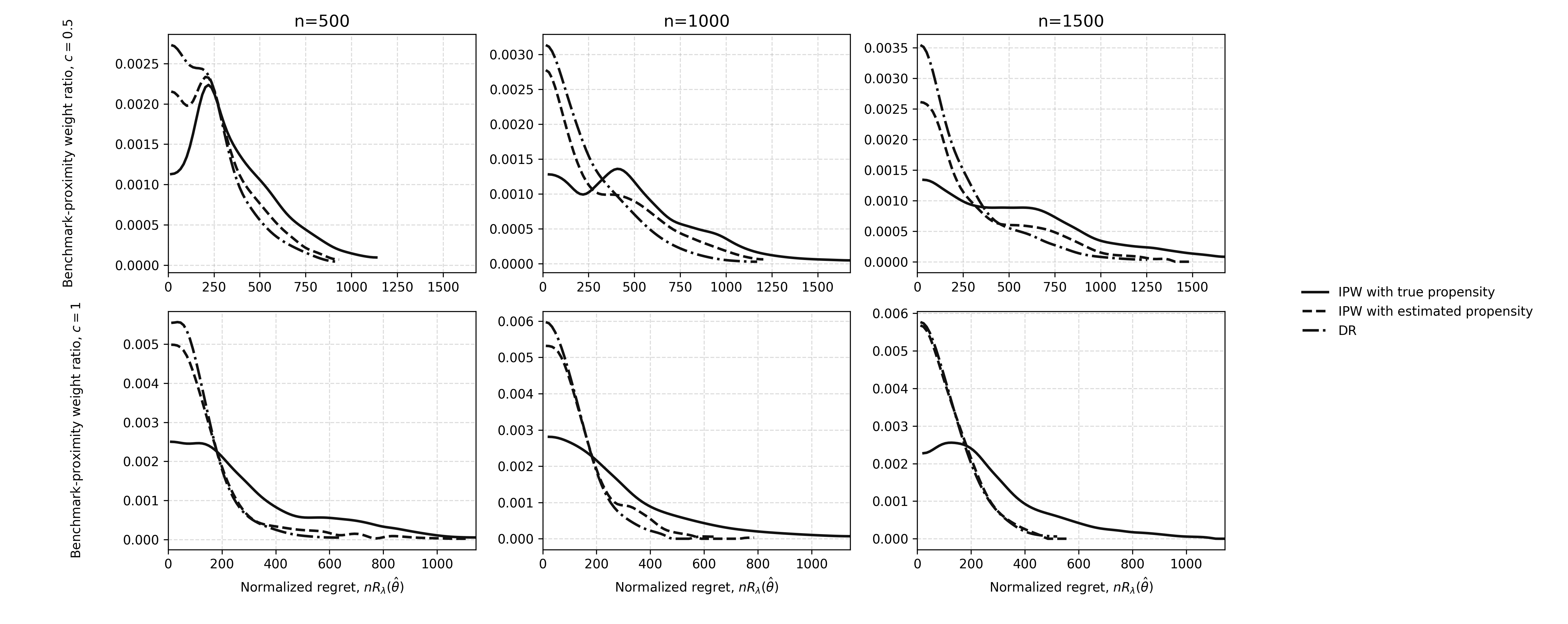}
    \caption{Density of normalized regret}
    \caption*{The figure reports estimated densities of normalized penalized-welfare regret, \(nR_\lambda(\hat\theta)\), across 1{,}000 Monte Carlo replications. Columns correspond to sample sizes and rows correspond to benchmark-proximity weight ratios \(c=\lambda/s_W\). The solid, dashed, and dotted curves correspond to true-propensity IPW, estimated-propensity IPW, and doubly robust estimation, respectively. Densities are estimated using \texttt{lpdensity} \citep{cattaneo2022lpdensity}, following \citet{cattaneo2020simple,cattaneo2024local}.}
    \label{fig:regret_density}
\end{figure}

Following Section~\ref{subsec:choosing-lambda}, we report
\(\lambda\) through \(c=\lambda/s_W\), where
\(s_W=\operatorname{sd}\{m_1^*(X)-m_0^*(X)\}\). In our implementation
\(s_W=2.728\), so \(c\in\{0.5,1\}\) corresponds to
\(\lambda\in\{1.364,2.728\}\). Each replication computes
\(\hat\theta^{tp}\), \(\hat\theta^{ep}\), and \(\hat\theta^{dr}\). Regret is evaluated on an independent oracle sample of size \(10^6\), using
\(R_\lambda(\hat\theta)=W_\lambda(\theta^*)-W_\lambda(\hat\theta)\), with
\(\theta^*=\arg\max_{\theta\in\Theta}W_\lambda(\theta)\). Table~\ref{tab:sim_result_binary} reports raw regret, and
Figure~\ref{fig:regret_density} plots normalized regret.

The results follow the predicted pattern. At every \((c,n)\), the efficient
criteria have lower mean penalized-welfare regret than true-propensity IPW. At
\(n=1500\), switching from true-propensity IPW to estimated-propensity IPW or
DR reduces mean regret by about 41 and 55 percent at \(c=0.5\), and by 56 and
57 percent at \(c=1\). At \(c=1\), the two efficient criteria nearly coincide
by \(n=1500\), as their common efficient limit predicts. In the density plots,
the efficient criteria are shifted left and have thinner right tails than
true-propensity IPW.

Two finite-sample features are useful for interpretation. First, regret is
larger at \(c=0.5\), where the weaker benchmark pull lets the oracle rule move
farther from the lottery and makes \(\theta^*\) harder to estimate. In this
more aggressive case, DR has lower regret than estimated-propensity IPW; this
is consistent with a more demanding balancing approximation and may also
reflect the random-forest calibration. Second, the \(c=0.5\) density plots show
that the efficient criteria stabilize between \(n=1000\) and \(n=1500\), while
the true-propensity density remains shifted to the right, consistent with the
additional assignment-noise component.

\section{Empirical Application}\label{sec:empirical}

The simulations provide a controlled comparison of regret. The empirical
application has a different role: it illustrates how to use the framework in a
real experiment. A single sample cannot rank estimators by realized welfare.
Instead, the application shows how to define the benchmark rule, choose an
interpretable assignment class, report the proximity preference, trace the
welfare--divergence frontier, and describe the implied reassignment.

We apply the methods of Section~\ref{sec:welfare_est} to the field experiment
of \citet{ashraf2006tying}, conducted with a rural bank in the Philippines to
evaluate a voluntary commitment savings product. The treatment has three arms:
no visit, a visit offering a commitment savings account, and a marketing visit
promoting standard savings accounts. The experimental assignment probabilities define the benchmark rule \(\pi^{\mathrm b}=(0.25,\,0.50,\,0.25)\),

corresponding to control, commitment, and marketing.

The outcome is the twelve-month change in total savings balance at the partner
bank, in pesos, winsorized at the 1st and 99th percentiles. We interpret this
as a savings objective, not a complete measure of client welfare: the
commitment product may also impose liquidity costs. The estimated rules should
therefore be read as improvements in measured savings subject to controlled
proximity to the experimental rule. After restricting log income per capita to
the 2.5th--97.5th percentiles, the analysis sample contains \(n=1{,}668\)
individuals: 444 control, 785 commitment, and 439 marketing.

The candidate rule is a benchmark-centered softmax with the control arm
normalized to zero,
\[
    \pi_{\theta}(t\mid x)
    =
    \frac{\pi^{\mathrm b}(t)\exp(\theta_t^\top x)}
         {\sum_{s=0}^{2}\pi^{\mathrm b}(s)\exp(\theta_s^\top x)},
    \qquad
    \theta_0\equiv0 .
\]
Thus \(\theta=0\) reproduces the experimental rule. The primary specification
uses log income per capita and recent account activity; alternative
specifications are reported in the Appendix.

\begin{figure}[p]
\centering
\includegraphics[width=0.78\linewidth]{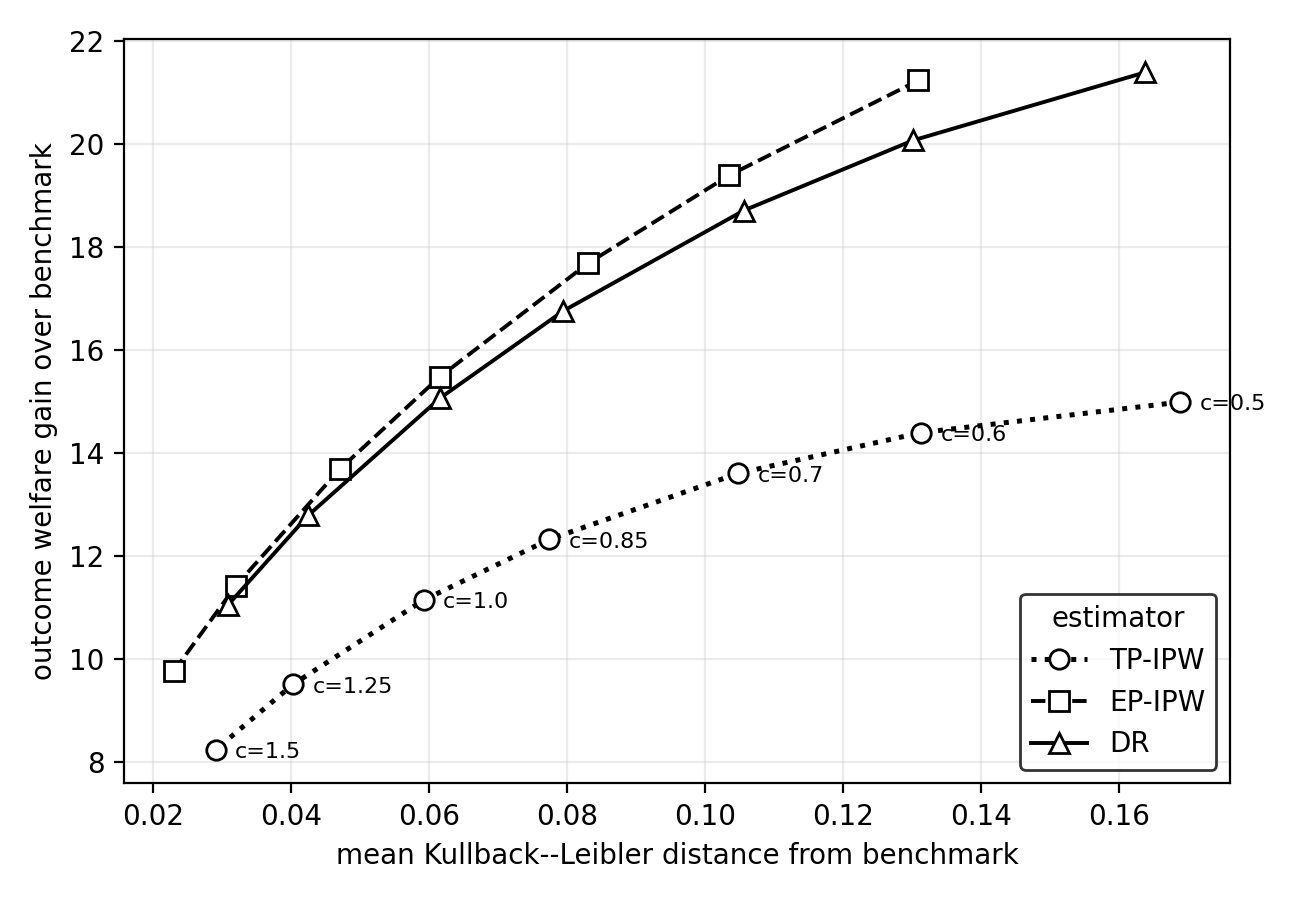}
\caption{Welfare--divergence frontier}
\caption*{Each point is a benchmark-centered softmax rule estimated at one value of the preference parameter \(c\); the vertical axis is the twelve-month savings gain over the experimental rule, and the horizontal axis is the average KL divergence from that rule. Labels on the TP-IPW curve indicate the corresponding value of \(c\); the marker at \(c = 1.0\) highlights the illustrative preference value used in the main-text exhibits.}
\label{fig:aky_frontier}
\end{figure}

We set \(\lambda=c\hat s_W\), where \(\hat s_W=157.7\) pesos is the
benchmark-weighted cross-arm dispersion of the cross-fitted outcome
regressions. Under the unrestricted tilt, an outcome contrast of size
\(\hat s_W\) changes the relative log assignment weight by \(1/c\). The main
tables use \(c=1.0\), a unit-scale reference preference rather than a
data-selected optimum. The frontier in Figure~\ref{fig:aky_frontier} shows the
tradeoff over the prespecified grid.

Nuisance functions are constructed by five-fold cross-fitting. The balancing
weights use the entropy-tilting procedure of Section~\ref{subsec:ipw_ep} with
a linear basis in female, recent account activity, log income per capita,
savings balance, and age. The outcome regression is a shallow random forest,
fit separately by arm, with \texttt{max\_depth}=3 and
\texttt{min\_samples\_leaf}=20. Every estimated rule is evaluated with the same
cross-fitted doubly robust evaluator, so differences across estimators reflect
the estimated rules rather than the scoring rule.

Table~\ref{tab:aky_policy_values} reports the benchmark and the three estimated
rules at \(c=1.0\). The table summarizes the scale of the rules, not a
single-sample ranking of estimators. All three rules have positive estimated
outcome gains over the benchmark. In this sample, the DR rule has the largest
unpenalized gain, while EP-IPW has the largest penalized value because it
attains a similar gain with a smaller KL departure. For the DR rule, the
estimated savings gain is 15.1 pesos, the average total-variation distance is
0.14, and the 95th-percentile probability shift is 0.20. Thus the estimated
rule makes moderate randomized adjustments to the experimental rule rather than
approaching a deterministic assignment. The bootstrap standard errors are large
relative to the gains, so we do not interpret the realized ordering as evidence
that one estimator dominates another in this application.

Table~\ref{tab:aky_coefficients} reports the softmax coefficients. Each
coefficient is a log-odds tilt for the corresponding arm relative to control.
The estimates are imprecise, but their signs describe the estimated rule. Log
income per capita has positive coefficients on both active-visit arms, so
higher-income clients are tilted away from the no-visit control. Recent account
activity has a negative coefficient on the commitment arm and a smaller, less
systematic coefficient on the marketing arm. Thus clients without recent
activity are tilted more strongly toward the commitment offer.

Figure~\ref{fig:aky_frontier} reports the welfare--divergence frontier for the
primary specification. Since \(c\) is a preference parameter, we do not select
it from the data. The frontier shows the menu of estimated rules: moving left
keeps the rule closer to the lottery and reduces estimated gains, while moving
right permits larger departures and larger estimated gains. The point
\(c=1.0\) is the unit-scale reference used in the main tables; the Appendix
reports frontiers for two alternative specifications.

The comparison across estimators is descriptive. EP-IPW generally lies above
and to the left of TP-IPW over the displayed range, and DR tracks EP-IPW
closely at moderate values of \(c\). This pattern is consistent with the HIR
mechanism in Section~\ref{sec:welfare_est}: even when assignment probabilities
are known by design, estimated balancing weights can remove
treatment-assignment noise left by true-propensity weighting. It is not a
single-sample welfare ranking.

\begin{figure}[p]
\centering
\includegraphics[width=0.78\linewidth]{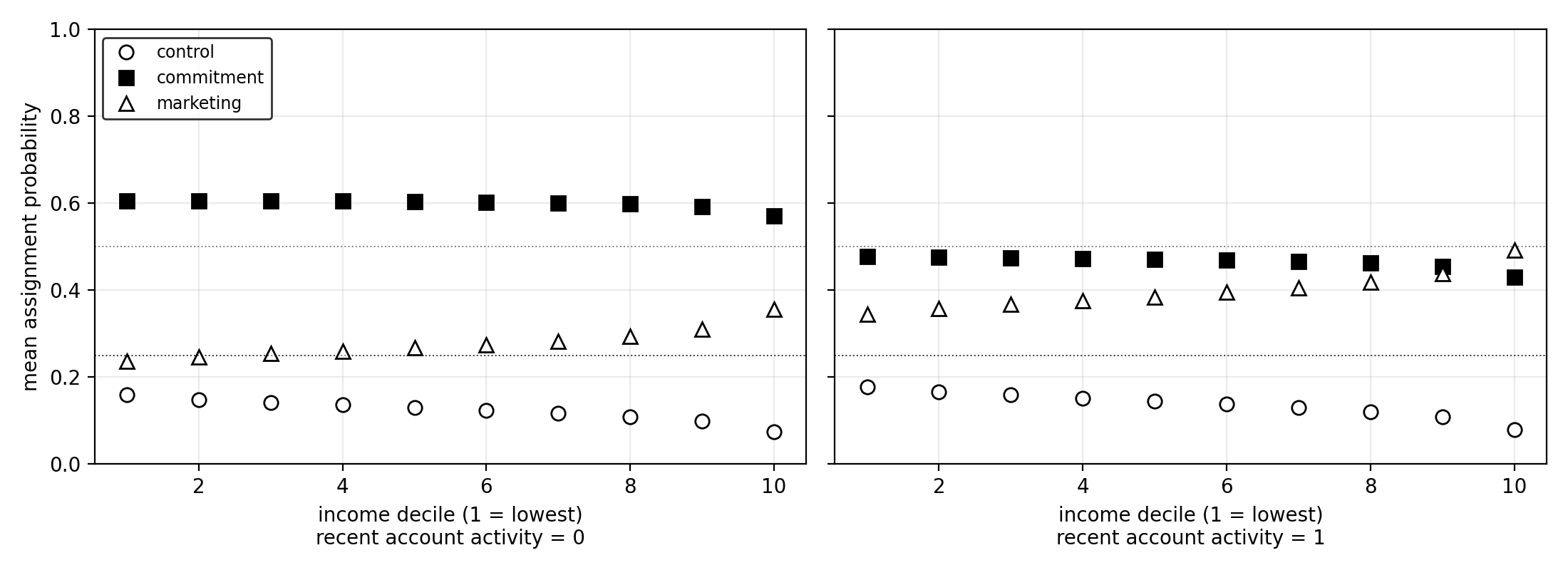}
\caption{DR rule assignment probabilities by income decile}
\caption*{At the illustrative preference value \(c = 1.0\). Dotted reference lines mark the benchmark probabilities \(\pi^{\mathrm b} = (0.25,\,0.50,\,0.25)\).}
\label{fig:aky_probs_by_income_decile}
\end{figure}

Figure~\ref{fig:aky_probs_by_income_decile} shows the DR assignment probabilities at \(c=1.0\), separately by recent account activity. In both groups, the rule gradually shifts higher-income clients away from control and toward marketing. Among clients without recent activity, the commitment arm remains dominant, with assignment probability near 0.60 across deciles, while control falls and marketing rises. Among clients with recent activity, the shift toward marketing is stronger: marketing increases across income deciles and slightly exceeds commitment in the top decile. Thus the learned rule makes moderate probability adjustments rather than introducing a sharp threshold.

\section{Conclusion}\label{sec:conclude}

We have studied efficient estimation of treatment assignment rules that improve outcome welfare while limiting departure from a known benchmark rule, such as a lottery, a status-quo mechanism, or an experimental design. The benchmark-penalized criterion gives randomized assignment an economic role and defines regret relative to the best rule in a specified class.

Two findings carry the main message. First, penalized-welfare regret has a second-order limiting distribution, so the mean of limiting regret is governed by the covariance of the assignment-rule estimator. An estimator that wastes first-order information delivers a rule whose expected performance is systematically, not merely noisily, worse. Second, an assignment-rule analogue of the Hirano--Imbens--Ridder phenomenon holds: true-propensity IPW is generally inefficient, while sufficiently rich estimated-propensity IPW and doubly robust estimation attain the efficient regret distribution. The same efficiency force reappears in welfare-process bounds for the unpenalized
deterministic case, so it is not an artifact of the smooth penalized criterion.

The practical recommendation runs against a natural instinct. Even in
randomized experiments, where assignment probabilities are known by design, weighting by the known propensity is not the precision benchmark. Under the conditions studied here, estimated-propensity IPW and doubly robust estimation can deliver more precise assignment rules. The calibrated simulations
illustrate the regret prediction, the commitment-savings application
illustrates the precision gains, and in both cases the estimated rules control their departure from the existing randomized mechanism.

Several directions remain open. Our regularity theory targets the best rule in a fixed finite-dimensional class; characterizing efficient regret when the class grows with the sample, or when the benchmark-proximity weight is selected in a data-driven way, would extend the results toward more flexible estimation of assignment rules. The benchmark-proximity criterion may also be of independent interest beyond efficiency comparisons, as a way of formalizing incremental reform of existing allocation mechanisms.

\clearpage
\begin{table}[p]
\centering
\renewcommand{\arraystretch}{1.25}
\begin{tabular}{ccccc}
\toprule
\(c=\lambda/s_W\) & \(n\)
& \(R_\lambda(\hat\theta^{tp})\)
& \(R_\lambda(\hat\theta^{ep})\)
& \(R_\lambda(\hat\theta^{dr})\) \\
\midrule
\multirow{3}{*}{0.5}
& 500  & 0.730 (0.477) & 0.542 (0.399) & 0.468 (0.365) \\
& 1000 & 0.492 (0.364) & 0.336 (0.285) & 0.255 (0.226) \\
& 1500 & 0.378 (0.293) & 0.222 (0.217) & 0.169 (0.164) \\
\addlinespace
\multirow{3}{*}{1.00}
& 500  & 0.610 (0.506) & 0.337 (0.376) & 0.238 (0.210) \\
& 1000 & 0.288 (0.263) & 0.135 (0.132) & 0.122 (0.107) \\
& 1500 & 0.183 (0.156) & 0.081 (0.067) & 0.078 (0.064) \\
\bottomrule
\end{tabular}
\caption{Penalized-welfare regret in the calibrated simulation}
\caption*{\footnotesize The table reports the mean and standard deviation, in parentheses, of penalized-welfare regret \(R_\lambda(\hat\theta)\) across 1{,}000 Monte Carlo replications. The three estimator columns correspond to true-propensity IPW, estimated-propensity IPW, and doubly robust estimation. The benchmark-proximity weight is reported as \(c=\lambda/s_W\), where \(s_W\) is the standard deviation of the oracle conditional treatment effect in the simulation DGP.}
\label{tab:sim_result_binary}
\end{table}

\clearpage

\clearpage
\begin{table}[p]
\centering
\renewcommand{\arraystretch}{1.20}
\begin{tabular}{lcccccc}
\toprule
rule & outcome & gain & mean KL & avg TV & p95 prob.\ shift & penalized \\
\midrule
Benchmark  & 68.29 &  0.00 & 0.000 & 0.000 & 0.000 & 68.29 \\
TP-IPW    & 79.44 & 11.15 & 0.059 & 0.137 & 0.187 & 70.11 \\
                  & (15.81) & (14.16) &   &      &      & (27.18) \\
EP-IPW    & 81.99 & 13.69 & 0.047 & 0.118 & 0.185 & 74.56 \\
                  & (16.41) & (14.77) &   &      &      & (24.24) \\
DR       & 83.36 & 15.07 & 0.062 & 0.135 & 0.195 & 73.65 \\
                  & (16.42) & (14.78) &   &      &      & (23.57) \\
\bottomrule
\end{tabular}
\caption{Policy values and distance from the experimental benchmark rule}
\caption*{At the illustrative preference value \(c = 1.0\) (primary specification, log income per capita + recent account activity). Common DR evaluator; bootstrap SEs in parentheses (\(B = 1{,}000\)). Outcomes are reported in pesos.}
\label{tab:aky_policy_values}
\end{table}

\clearpage
\begin{table}[p]
\centering
\renewcommand{\arraystretch}{1.20}
\begin{tabular}{llccc}
\toprule
treatment & covariate & TP-IPW & EP-IPW & DR \\
\midrule
commitment  & log income per capita & 0.320  & 0.232  & 0.215 \\
                   &                       & (0.505) & (0.485) & (0.515) \\
                   & recent account activity & $-$0.970 & $-$0.454 & $-$0.355 \\
                   &                         & (1.237) & (1.188) & (1.248) \\
                   & constant              & 1.099  & 0.869  & 0.970 \\
                   &                       & (0.560) & (0.578) & (0.561) \\
\midrule
marketing   & log income per capita & 0.334  & 0.371  & 0.350 \\
                   &                       & (0.863) & (1.037) & (0.946) \\
                   & recent account activity & $-$0.262 & 0.063 & 0.258 \\
                   &                         & (1.658) & (1.634) & (1.643) \\
                   & constant              & 1.035  & 0.857  & 0.927 \\
                   &                       & (0.785) & (0.802) & (0.692) \\
\bottomrule
\end{tabular}
\caption{Policy coefficients for the benchmark-centered softmax rule}
\caption*{At the illustrative preference value \(c = 1.0\), primary specification. Robust bootstrap standard errors (\(\mathrm{MAD}\times 1.4826\), \(B = 1{,}000\)) in parentheses. Each coefficient is a log-odds tilt away from the experimental rule \(\pi^{\mathrm b}\) on the indicated arm.}
\label{tab:aky_coefficients}
\end{table}

\clearpage

\clearpage

\clearpage

\bibliographystyle{chicago}
\bibliography{references.bib}

\newpage
\appendix 
\begin{center}
{\large\bf APPENDIX}
\end{center}
\numberwithin{theorem}{section}
\numberwithin{lemma}{section}
\numberwithin{equation}{section}
\numberwithin{proposition}{section}
\numberwithin{assumption}{section}

\section{Regularity details for deterministic assignment rules}
\label{app:regularity-details}

We begin by recalling the notion of pathwise differentiability. Let
$\mathcal P$ denote the statistical model satisfying Assumptions 1 and 2,
with the benchmark assignment rule $\Pi^{\mathrm b}$ and the assignment class
$\{\Pi_\theta:\theta\in\Theta\}$ treated as fixed. For a given
$P\in\mathcal P$, a regular parametric submodel through $P$ is a collection
$\{P_\varepsilon:\varepsilon\in(-r,r)\}\subset\mathcal P$ such that
$P_0=P$, all $P_\varepsilon$ are dominated by a common $\sigma$-finite
measure, and the map
$\varepsilon\mapsto \log\frac{dP_\varepsilon}{d\bar\nu}(Z)$ is
differentiable at $0$ in $L^2(P)$ for some dominating measure $\bar\nu$.
The score of the path is
$s(Z)=\frac{\partial}{\partial\varepsilon}
\log\frac{dP_\varepsilon}{d\bar\nu}(Z)\big|_{\varepsilon=0}$ in
$L_0^2(P)$, where
$L_0^2(P)=\{h\in L^2(P):\mathbb E_P[h(Z)]=0\}$. The tangent space
$\mathcal S$ is the $L^2(P)$-closure of all such scores.

A parameter $\psi:\mathcal P\to\mathbb R^d$ is pathwise differentiable at
$P$ relative to $\mathcal S$ if there exists
$\phi\in\mathcal S\subset L_0^2(P)$ such that, for every regular submodel
with score $s$,
\[
    \frac{\partial}{\partial\varepsilon}\psi(P_\varepsilon)
    \bigg|_{\varepsilon=0}
    =\mathbb E_P[\phi(Z)s(Z)].
\]
The function $\phi$ is the influence function of $\psi$. Pathwise
differentiability is necessary for the existence of $\sqrt n$-regular
estimators of $\psi$ in $\mathcal P$
\citep{van1991differentiable, van1998asymptotic}.\footnote{An estimator is
regular at the true law $P$ if its first-order limiting distribution is stable
under all $1/\sqrt n$-local, contiguous perturbations of $P$.}

The main target of this paper is the best rule in a smooth randomized
assignment class for the penalized welfare criterion $W_{\lambda}$. The
smoothness of the assignment rule is important for regular estimation. To
clarify the contrast with deterministic assignment, we first discuss the
pathwise differentiability of the values of deterministic rules.

The source of nonregularity differs across treatment spaces. With binary or
finite discrete treatments, the value of a fixed deterministic rule can be a
regular treatment-effect functional under standard overlap conditions. The
nonregularity in deterministic treatment choice instead comes from learning a
hard assignment rule, such as an argmax or threshold rule, whose indicator
structure is nonsmooth. This is the source of cube-root-type behavior in
threshold treatment-choice problems; see, for example, \citet{crippa2025regret}
and the cube-root asymptotics of \citet{kim1990cube}.

With continuous treatments, the issue is sharper: even the value of a fixed
deterministic rule can fail to be pathwise differentiable. Consider a
deterministic rule $\Pi_\theta(dt|x)=\delta_{g_\theta(x)}(dt)$, where
$\delta_a$ denotes the Dirac measure at $a$. Its outcome-welfare component is
$W_0(\theta)=\mathbb E[m(g_\theta(X),X)]$. Although this expression resembles
an average of a conditional mean, the conditional mean $m(t,x)$ is evaluated
only along the graph $\{(g_\theta(x),x):x\in\mathcal X\}$, which is a
lower-dimensional subset of the joint $(T,X)$ support when $T$ is continuous.

Throughout this section, fix a deterministic class
$\Pi_\theta(dt|x)=\delta_{g_\theta(x)}(dt)$ and suppose the conditional law of
$T|X=x$ has a Lebesgue density $f(\cdot|x)$ for $P_X$-almost every $x$. For a
parameter value $\theta$, a measurable set $A\subseteq\mathcal X$, and
$\delta>0$, write
\[
    \mathcal N_\theta(A,\delta)
    =\{(t,x):x\in A,\ |t-g_\theta(x)|\le\delta\}
\]
for the $\delta$-neighborhood of the graph of $g_\theta$ over $A$. The first
two assumptions are stated for a generic parameter value $\theta$: Theorem
\ref{thm:deterministic_welfare_irregular} uses them at the fixed $\theta$
under consideration, and Proposition~\ref{prop:theta_star_irregular} uses
them at $\theta=\theta^*$.

\begin{assumption}
\label{ass:local-nondegeneracy}
There exist a measurable set $A\subseteq\mathcal X$ with $P_X(A)>0$, a number $\delta>0$, and constants $0<\underline f\le\bar f<\infty$ and $\underline\sigma^2>0$ such that \(\underline f\le f(t|x)\le\bar f\) and \(\underline\sigma^2\le \sigma^2(t,x)<\infty\) for all \((t,x)\in\mathcal N_\theta(A,\delta)\).

\end{assumption}

% In particular, $g_\theta(x)$ lies in the support of $T|X=x$ for every $x\in A$, so the graph is reachable by the data.

\begin{assumption}\label{ass:mean-perturbations}
Let $\mathcal H_\theta$ denote the set of bounded measurable functions
$h:\mathcal T\times\mathcal X\to\mathbb R$ that are supported in
$\mathcal N_\theta(A,\delta)$ and continuously differentiable in $t$. For every
$h\in\mathcal H_\theta$, the model contains a regular submodel
$\{P_{\varepsilon,h}\}$ through $P$ that leaves the law of $(T,X)$ fixed and
satisfies
\[
    s_h(Z)=\frac{Y-m(T,X)}{\sigma^2(T,X)}h(T,X),
    \qquad
    \left.\frac{\partial}{\partial\varepsilon}
    m_{P_{\varepsilon,h}}(t,x)\right|_{\varepsilon=0}=h(t,x).
\]
\end{assumption}

A path with score $s_h$ that fixes the law of $(T,X)$ has conditional-mean
derivative $\mathbb E[Ys_h(Z)|T=t,X=x]=h(t,x)$, so the substantive requirement
in Assumption~\ref{ass:mean-perturbations} is that the model contains regular
paths with these scores.

\begin{theorem}
\label{thm:deterministic_welfare_irregular}
Fix $\theta$ and suppose Assumptions~\ref{ass:local-nondegeneracy} and
\ref{ass:mean-perturbations} hold at $\theta$. Then the deterministic-rule outcome-welfare functional \(W_0(\theta)=\mathbb E[m(g_\theta(X),X)]\) is not pathwise differentiable at $P$.
\end{theorem}

The theorem is a continuous-treatment result. For finite discrete treatments,
the value of a fixed deterministic rule is typically regular under overlap; the
nonregularity in deterministic treatment choice then comes from the hard argmax
or threshold map used to learn the rule. With continuous treatments, even the
value of a fixed deterministic rule requires point evaluation of the conditional
mean along a zero-measure graph.

\begin{proof}
Suppose, toward a contradiction, that $W_0(\theta)$ is pathwise
differentiable. Then there exists $\varphi_\theta\in L_0^2(P)$ such that, for
every regular submodel with score $s$,
\begin{align}
\label{eq:det-if-representation}
    \left.\frac{\partial}{\partial\varepsilon}
    W_{0,P_\varepsilon}(\theta)\right|_{\varepsilon=0}
    =\mathbb E[\varphi_\theta(Z)s(Z)].
\end{align}

Apply this representation to the submodel of
Assumption~\ref{ass:mean-perturbations} indexed by
$h\in\mathcal H_\theta$. Since the law of $(T,X)$ is fixed and the derivative
of the conditional mean is $h$,
\[
    \left.\frac{\partial}{\partial\varepsilon}
    W_{0,P_{\varepsilon,h}}(\theta)\right|_{\varepsilon=0}
    =\mathbb E[h(g_\theta(X),X)].
\]
Define
\[
    \eta_\theta(t,x)
    =\mathbb E\left[
        \varphi_\theta(Z)
        \frac{Y-m(T,X)}{\sigma^2(T,X)}
        \Bigm|T=t,X=x
    \right].
\]
By Cauchy--Schwarz and the variance bound in
Assumption~\ref{ass:local-nondegeneracy}, the restriction of $\eta_\theta$ to
$\mathcal N_\theta(A,\delta)$ is square integrable:
\[
    \int_{\mathcal N_\theta(A,\delta)}
    \eta_\theta^2\,dP_{T,X}
    \leq
    \frac{\mathbb E[\varphi_\theta(Z)^2]}{\underline\sigma^2}
    <\infty .
\]
Combining the first two displays gives
\begin{align}
\label{eq:point-eval-representation}
    \mathbb E[h(g_\theta(X),X)]
    =\mathbb E[\eta_\theta(T,X)h(T,X)]
\end{align}
for every $h\in\mathcal H_\theta$.

Choose $\kappa\in C_c^\infty((-1,1))$ with $\kappa(0)\neq0$, and for
$n\geq1/\delta$ define
\[
    h_n(t,x)=n^{1/2}
    \kappa\{n(t-g_\theta(x))\}
    \mathbf 1\{x\in A\}.
\]
Each $h_n$ is bounded, continuously differentiable in $t$, and supported in
$\mathcal N_\theta(A,\delta)$, hence belongs to $\mathcal H_\theta$. Their
$L_2(P_{T,X})$ norms are uniformly bounded because
\[
\begin{aligned}
    \mathbb E[h_n(T,X)^2]
    & =\mathbb E\left[
        \mathbf 1\{X\in A\}
        \int n\kappa\{n(t-g_\theta(X))\}^2 f(t|X)dt
    \right]  \\
    & =\mathbb E\left[
        \mathbf 1\{X\in A\}
        \int \kappa(u)^2 f(g_\theta(X)+u/n|X)du
    \right]
    \leq \bar f P_X(A)\int \kappa(u)^2du .
\end{aligned}
\]
On the other hand,
\[
    \mathbb E[h_n(g_\theta(X),X)]
    =n^{1/2}\kappa(0)P_X(A),
\]
which diverges in absolute value. This contradicts
\eqref{eq:point-eval-representation}: since $h_n$ is supported in
$\mathcal N_\theta(A,\delta)$, its right-hand side is bounded by
\[
    \left(\int_{\mathcal N_\theta(A,\delta)}
    \eta_\theta^2\,dP_{T,X}\right)^{1/2}
    \|h_n\|_{L_2(P_{T,X})}.
\]
Thus no square-integrable influence function can represent the pathwise
derivative of $W_0(\theta)$, and the deterministic-rule value is not pathwise
differentiable.
\end{proof}

The corresponding result for the best deterministic rule requires two further
conditions: a regular interior argmax, and an expansion of the welfare gradient
along the perturbations of Assumption~\ref{ass:mean-perturbations}.

\begin{assumption}\label{ass:argmax-regularity}
$\theta^*$ is an interior local maximizer of $W_0$, there is an open neighborhood $N\subseteq\operatorname{int}(\Theta)$ of $\theta^*$ on which $W_0$ is continuously differentiable, and the gradient of $W_0$ is differentiable at $\theta^*$ with \(H=-\frac{\partial^2 W_0(\theta)}{\partial\theta\partial\theta'} \big|_{\theta=\theta^*}\) nonsingular. For $P_X$-almost every $x$, the map $\theta\mapsto g_\theta(x)$ is continuously differentiable on $N$. In addition, there exist a nonzero vector $a\in\mathbb R^p$ and a constant $c_0>0$ such that, with $A$ as in Assumption~\ref{ass:local-nondegeneracy} at $\theta^*$, \(\left|a'H^{-1}\frac{\partial g_{\theta^*}(x)}{\partial\theta}\right|\geq c_0\) {for all } \(x\in A\).
\end{assumption}

\begin{assumption}\label{ass:welfare-expansion}
For every $h\in\mathcal H_{\theta^*}$, let $\{P_{\varepsilon,h}\}$ be the
submodel in Assumption~\ref{ass:mean-perturbations}, and write
$L_h(\theta)=\mathbb E[h(g_\theta(X),X)]$. Then:
\begin{enumerate}[label=(\roman*)]
\item uniformly for $\theta\in N$,
\[
    \frac{\partial W_{0,P_{\varepsilon,h}}(\theta)}{\partial\theta}
    =\frac{\partial W_{0,P}(\theta)}{\partial\theta}
    +\varepsilon\,\frac{\partial L_h(\theta)}{\partial\theta}
    +o(\varepsilon);
\]
\item the map $\theta\mapsto\partial L_h(\theta)/\partial\theta$ is continuous
at $\theta^*$, and
\[
    \frac{\partial L_h(\theta^*)}{\partial\theta}
    =\mathbb E\left[
        \frac{\partial h(g_{\theta^*}(X),X)}{\partial t}
        \frac{\partial g_{\theta^*}(X)}{\partial\theta}
    \right];
\]
\item $W_{0,P_{\varepsilon,h}}$ has a unique maximizer
$\theta^*(P_{\varepsilon,h})$ over $\Theta$, which lies in $N$, for all
sufficiently small $\varepsilon$.
\end{enumerate}
\end{assumption}

\begin{proposition}
\label{prop:theta_star_irregular}
Suppose Assumptions~\ref{ass:local-nondegeneracy} and
\ref{ass:mean-perturbations} hold at $\theta=\theta^*$, and
Assumptions~\ref{ass:argmax-regularity} and \ref{ass:welfare-expansion} hold.
Then the argmax map $P\mapsto\theta^*(P)$ is not pathwise differentiable at
$P$.
\end{proposition}

\begin{proof}
Suppose, toward a contradiction, that $P\mapsto\theta^*(P)$ is pathwise
differentiable. Fix $h\in\mathcal H_{\theta^*}$ and write
$\theta_{\varepsilon,h}=\theta^*(P_{\varepsilon,h})$, which is well defined and
interior for all sufficiently small $\varepsilon$ by
Assumption~\ref{ass:welfare-expansion}(iii), so that
\[
    \frac{\partial W_{0,P_{\varepsilon,h}}
    (\theta_{\varepsilon,h})}{\partial\theta}=0,
    \qquad
    \frac{\partial W_{0,P}(\theta^*)}{\partial\theta}=0 .
\]
Pathwise differentiability implies
$\theta_{\varepsilon,h}-\theta^*=O(\varepsilon)$. Evaluating the expansion in
Assumption~\ref{ass:welfare-expansion}(i) at $\theta_{\varepsilon,h}$,
expanding
$\partial W_{0,P}(\theta_{\varepsilon,h})/\partial\theta
=-H(\theta_{\varepsilon,h}-\theta^*)
+o(\|\theta_{\varepsilon,h}-\theta^*\|)$ around $\theta^*$, and using the
continuity in Assumption~\ref{ass:welfare-expansion}(ii) to replace
$\partial L_h(\theta_{\varepsilon,h})/\partial\theta$ by
$\partial L_h(\theta^*)/\partial\theta+o(1)$, gives
\[
    0=-H(\theta_{\varepsilon,h}-\theta^*)
      +\varepsilon
      \frac{\partial L_h(\theta^*)}{\partial\theta}
      +o(\varepsilon).
\]
Hence
\begin{align}
\label{eq:argmax-directional-derivative}
    \left.
    \frac{\partial \theta^*(P_{\varepsilon,h})}{\partial\varepsilon}
    \right|_{\varepsilon=0}
    =H^{-1}
    \frac{\partial L_h(\theta^*)}{\partial\theta}.
\end{align}

The scalar map $P\mapsto a'\theta^*(P)$ would then also be pathwise
differentiable. Hence there exists $\psi_a\in L_0^2(P)$ such that, for every
$h\in\mathcal H_{\theta^*}$,
\[
    a'
    \left.
    \frac{\partial \theta^*(P_{\varepsilon,h})}{\partial\varepsilon}
    \right|_{\varepsilon=0}
    =\mathbb E[\psi_a(Z)s_h(Z)].
\]
Define
\[
    \eta_a(t,x)=\mathbb E\left[
        \psi_a(Z)
        \frac{Y-m(T,X)}{\sigma^2(T,X)}
        \Bigm|T=t,X=x
    \right].
\]
As in the proof of Theorem~\ref{thm:deterministic_welfare_irregular}, the
restriction of $\eta_a$ to $\mathcal N_{\theta^*}(A,\delta)$ is square
integrable, which suffices below because every $h\in\mathcal H_{\theta^*}$ is
supported there. Combining the last display with
\eqref{eq:argmax-directional-derivative} and the interchange in
Assumption~\ref{ass:welfare-expansion}(ii) yields
\begin{align}
\label{eq:argmax-unbounded-functional}
    \mathbb E[\eta_a(T,X)h(T,X)]
    =\mathbb E\left[
        \frac{\partial h(g_{\theta^*}(X),X)}{\partial t}
        q(X)
    \right],
    \qquad
    q(x)=a'H^{-1}
    \frac{\partial g_{\theta^*}(x)}{\partial\theta} .
\end{align}

Choose $\kappa\in C_c^\infty((-1,1))$ with $\kappa'(0)\neq0$, and for
$n\geq1/\delta$ define
\[
    h_n(t,x)=n^{1/2}
    \kappa\{n(t-g_{\theta^*}(x))\}
    \operatorname{sgn}\{q(x)\}
    \mathbf 1\{x\in A\},
\]
which belongs to $\mathcal H_{\theta^*}$. As in the proof of
Theorem~\ref{thm:deterministic_welfare_irregular},
$\sup_n\|h_n\|_{L_2(P_{T,X})}<\infty$. But
\[
    \frac{\partial h_n(g_{\theta^*}(x),x)}{\partial t}
    =n^{3/2}\kappa'(0)
    \operatorname{sgn}\{q(x)\}
    \mathbf 1\{x\in A\},
\]
so the right-hand side of \eqref{eq:argmax-unbounded-functional} equals \(n^{3/2}\kappa'(0)\mathbb E[|q(X)|\mathbf 1\{X\in A\}]\),
which diverges in absolute value because
$\mathbb E[|q(X)|\mathbf 1\{X\in A\}]\geq c_0P_X(A)>0$. The left-hand side of
\eqref{eq:argmax-unbounded-functional} is bounded by the restricted $L_2$ norm
of $\eta_a$ times $\|h_n\|_{L_2(P_{T,X})}$, a contradiction. Therefore the
argmax map cannot be pathwise differentiable.
\end{proof}

Proposition~\ref{prop:theta_star_irregular} gives the corresponding argmax
result. Under smooth local uniqueness conditions for a continuous-treatment
deterministic class, the best deterministic rule inherits the same
point-evaluation nonregularity. This is why the regularity theory in the main
text is stated for smooth randomized rules and a fixed positive
benchmark-proximity weight.

\section{Primitive regularity for benchmark-centered exponential tilts}
\label{app:primitive-regularity}

This section gives simple sufficient conditions for Assumption~3 in the main text for the benchmark-centered exponential-tilt class. The conditions are not necessary; their purpose is only to show that the high-level interiority, uniqueness, and curvature requirements can be verified from the population criterion.

Consider the class
\[
    \Pi_\theta(dt|x)
    =
    \frac{\exp\{\theta'q(t,x)\}}
    {\int_{\mathcal T}\exp\{\theta'q(s,x)\}\Pi^{\mathrm b}(ds|x)}
    \Pi^{\mathrm b}(dt|x),
    \qquad \theta\in\Theta\subset\mathbb R^p.
\]
For the formulas below, covariance and variance are computed under the assignment distribution \(\Pi_\theta(\cdot|X)\), conditional on \(X\). The first assumption gives the differentiability needed for the closed-form gradient in Lemma~\ref{app-lem:tilt-gradient}. The second collects the curvature and boundary conditions used to verify Assumption~3 in the main text.

\begin{assumption}\label{ass:tilt-moments}
There is a bounded open set $\Theta'\supseteq\Theta$ such that, for
$P_X$-almost every $x$:
\begin{enumerate}[label=(\roman*)]
\item
\(\int_{\mathcal T}\{1+|m(s,x)|\}\exp\{\theta'q(s,x)\}\Pi^{\mathrm b}(ds|x)<\infty\) for all \(\theta\in\Theta'\);
\item
\(\mathbb E[\sup_{\theta\in\Theta'}
    \{\left\|\operatorname{Cov}_{\Pi_\theta}
    \{q(T,X),m(T,X)|X\}\right\| +
    \left\|\operatorname{Var}_{\Pi_\theta}
    \{q(T,X)|X\}\right\|\}]<\infty\).

\end{enumerate}
\end{assumption}

Condition (i), imposed on an open set, is the local exponential-moment condition that permits differentiating the conditional integrals defining \(\mu_\theta(x)\) and \(kl_\theta(x)\). Condition (ii) supplies the integrable envelope needed to differentiate the expectation over \(X\).

\begin{lemma}
\label{app-lem:tilt-gradient}
Under Assumption~\ref{ass:tilt-moments}, for every $\theta\in\Theta$,
\[
    \frac{\partial W_\lambda(\theta)}{\partial\theta}
    =
    \mathbb E\left[
        \operatorname{Cov}_{\Pi_\theta}
        \{q(T,X),m(T,X)|X\}
        -
        \lambda
        \operatorname{Var}_{\Pi_\theta}
        \{q(T,X)|X\}\theta
    \right].
\]
\end{lemma}

\begin{proof}
Fix $x$ and $\theta\in\Theta'$. By Assumption~\ref{ass:tilt-moments}(i) and the usual differentiability of Laplace-type integrals on open sets, applied to the positive and negative parts of the integrands, the maps
\[
\begin{aligned}
    &\theta\mapsto\int_{\mathcal T}
    \exp\{\theta'q(s,x)\}\,\Pi^{\mathrm b}(ds|x),  \\
    &\theta\mapsto\int_{\mathcal T}
    m(s,x)\exp\{\theta'q(s,x)\}\,\Pi^{\mathrm b}(ds|x)
\end{aligned}
\]
are differentiable, with derivatives obtained by differentiating under the integral sign. The log density of \(\Pi_\theta(\cdot|x)\) relative to the benchmark rule is
\[
    \theta'q(t,x)
    -
    \log\int_{\mathcal T}
        \exp\{\theta'q(s,x)\}\Pi^{\mathrm b}(ds|x).
\]
Differentiating this log density gives the conditional score
\[
    q(t,x)-\int_{\mathcal T}q(s,x)\Pi_\theta(ds|x).
\]
Therefore, differentiating the outcome component
\(\mu_\theta(x)=\int m(t,x)\Pi_\theta(dt|x)\) gives
\[
    \frac{\partial \mu_\theta(x)}{\partial\theta}
    =
    \operatorname{Cov}_{\Pi_\theta}
    \{q(T,X),m(T,X)|X=x\}.
\]
For the KL component,
\[
    kl_\theta(x)
    =\theta'\mathbb E_{\Pi_\theta}[q(T,X)|X=x]
    -
    \log\int_{\mathcal T}
    \exp\{\theta'q(s,x)\}\Pi^{\mathrm b}(ds|x).
\]
Since
\[
    \frac{\partial}{\partial\theta'}
    \mathbb E_{\Pi_\theta}[q(T,X)|X=x]
    =
    \operatorname{Var}_{\Pi_\theta}\{q(T,X)|X=x\}
\]
and the log normalizing constant has gradient
\(\mathbb E_{\Pi_\theta}[q(T,X)|X=x]\),
\[
    \frac{\partial kl_\theta(x)}{\partial\theta}
    =
    \operatorname{Var}_{\Pi_\theta}\{q(T,X)|X=x\}\theta .
\]
Finally, Assumption~\ref{ass:tilt-moments}(ii) provides an integrable envelope over $\Theta'$, so the expectation over $X$ may also be differentiated under the integral sign. Taking expectations over $X$ gives the displayed gradient.
\end{proof}

% The gradient has a useful interpretation. The covariance term tilts the rule toward treatments whose conditional mean outcomes covary positively with the assignment features. The second term is the pull back toward the benchmark generated by the KL penalty. Full-rank variation in the assignment features makes this pull nondegenerate. The remaining requirements concern the shape of the population criterion.

\begin{assumption}\label{ass:tilt-criterion}
$\Theta=\{\theta:\|\theta\|\le R\}$, the criterion \(W_\lambda\) is twice continuously differentiable on a neighborhood of \(\Theta\), and the following conditions hold for all \(\theta\in\Theta\):
% the assignment features have uniformly full-rank variation: 
(i) \(\lambda_{\min}\left(\mathbb E\left[\operatorname{Var}_{\Pi_\theta}\{q(T,X)|X\}\right]\right)\geq \underline\kappa>0\);
% the feature--outcome covariance term is uniformly bounded:
(ii) \(\left\|
        \mathbb E\left[
        \operatorname{Cov}_{\Pi_\theta}
        \{q(T,X),m(T,X)|X\}\right]
    \right\|
    \leq B<\infty\);
% the boundary is far enough from the origin:
(iii) \(\lambda\underline\kappa R>B\);
(iv) % the criterion is strongly concave:
\(-\frac{\partial^2 W_\lambda(\theta)}{\partial\theta\partial\theta'}\succeq \underline h I_p\) for some \(\underline h>0\).
\end{assumption}

\begin{proposition}
\label{app-prop:tilt-regularity}
Under Assumptions~\ref{ass:tilt-moments} and \ref{ass:tilt-criterion}, \(W_\lambda\) has a unique maximizer \(\theta^*\in\operatorname{int}(\Theta)\), the maximizer is separated, and \( H= -\frac{\partial^2 W_\lambda(\theta)}{\partial\theta\partial\theta'}\big|_{\theta=\theta^*}\) is positive definite. Hence Assumption~3 in the main text holds.
\end{proposition}

\begin{proof}
Continuity of \(W_\lambda\) on the compact set \(\Theta\) gives existence of a maximizer. We first rule out boundary maximizers. If \(\|\theta\|=R\), Lemma~\ref{app-lem:tilt-gradient} and Assumption~\ref{ass:tilt-criterion}(i)--(iii) imply
\[
    \theta'
    \frac{\partial W_\lambda(\theta)}{\partial\theta}
    \leq
    R B
    -
    \lambda\underline\kappa R^2
    <0.
\]
Thus moving slightly inward from any boundary point increases the criterion, so a maximizer cannot lie on the boundary.

Let \(\theta^*\) be an interior maximizer, so that
\(\partial W_\lambda(\theta^*)/\partial\theta=0\). For any \(\theta\in\Theta\), a second-order expansion along the segment from \(\theta^*\) to \(\theta\), which lies in \(\Theta\), together with Assumption~\ref{ass:tilt-criterion}(iv), gives
\[
    W_\lambda(\theta)
    \leq W_\lambda(\theta^*)
    -\frac{\underline h}{2}\|\theta-\theta^*\|^2 .
\]
Hence the maximizer is unique and separated. Assumption~\ref{ass:tilt-criterion}(iv) at \(\theta=\theta^*\) gives \(H\succeq \underline h I_p\), so the negative Hessian at the maximizer is positive definite. This verifies Assumption~3 in the main text.
\end{proof}

\section{Proofs of lemmas and theorems in the main text}
Throughout the proofs for the penalized-welfare results, fix $\lambda>0$ and write
\[
    kl_\theta(X)=\mathrm{KL}\{\Pi_\theta(\cdot|X)\|\Pi^{\mathrm b}(\cdot|X)\}.
\]
When derivatives with respect to $\theta$ are used, write
\[
    \dot\pi_\theta(t|x)=\frac{\partial\pi_\theta(t|x)}{\partial\theta},
    \qquad
    \dot\mu_\theta(x)=\frac{\partial\mu_\theta(x)}{\partial\theta},
    \qquad
    \dot{kl}_\theta(x)=\frac{\partial kl_\theta(x)}{\partial\theta}.
\]
To keep the arguments below concise, we impose the following regularity condition in this proof section.

\medskip
\noindent\textbf{Condition (S).}
For $\nu\otimes P_X$-almost every $(t,x)$, $\theta\mapsto\pi_\theta(t|x)$
is twice continuously differentiable on $\Theta$; for $P_X$-almost every $x$,
$\theta\mapsto kl_\theta(x)$ is twice continuously differentiable on $\Theta$.
Moreover, there is a constant $\bar\pi<\infty$ such that
\[
    \sup_{\theta\in\Theta}\sup_{t,x}
    \left\{
    \pi_\theta(t|x)+\|\dot\pi_\theta(t|x)\|
    +\left\|
    \frac{\partial^2\pi_\theta(t|x)}{\partial\theta\partial\theta'}
    \right\|
    \right\}
    \le\bar\pi,
\]
\[
    \sup_{\theta\in\Theta}\sup_x
    \left\{
    kl_\theta(x)+\|\dot{kl}_\theta(x)\|
    +\left\|
    \frac{\partial^2kl_\theta(x)}{\partial\theta\partial\theta'}
    \right\|
    \right\}
    \le\bar\pi,
\]
and
\[
    \sup_{\theta\in\Theta}\int_{\mathcal T}
    \left\{
    \|\dot\pi_\theta(t|x)\|
    +
    \left\|
    \frac{\partial^2\pi_\theta(t|x)}{\partial\theta\partial\theta'}
    \right\|
    \right\}
    d\nu(t)
    \le\bar\pi
\]
for $P_X$-almost every $x$.

\medskip
Condition (S) is satisfied by the leading benchmark-centered exponential-tilt
class under bounded assignment features, compact $\Theta$, and a bounded
benchmark density. In that case the first two derivatives of $\pi_\theta$ are
bounded by fixed multiples of the benchmark density, their $\nu$-integrals are
bounded because $\pi_\theta$ integrates to one, and $kl_\theta$ has bounded
first and second derivatives. Under Condition (S), Assumption 5, and
compactness of $\Theta$, the Glivenko--Cantelli and uniform-Hessian statements
used below follow from standard finite-dimensional uniform laws
\citep[e.g.][Example 19.8]{van1998asymptotic}.

\begin{proof}[Proof of Lemma 1]
Fix $\theta$ and consider any regular submodel $\{P_\varepsilon\}$ with score
$s(Z)$. Under the factorization
\[
    f_Z(z)=f_X(x)f_{T|X}(t|x)f_{Y|T,X}(y|t,x),
\]
the score decomposes as
\[
    s(Z)=s_X(X)+s_p(T,X)+s_Y(Y,T,X),
\]
where
\[
    \mathbb E[s_X(X)]=0,\qquad
    \mathbb E[s_p(T,X)|X]=0,\qquad
    \mathbb E[s_Y(Y,T,X)|T,X]=0.
\]

For fixed $\theta$,
\[
    W_\lambda(\theta)=\mathbb E[\mu_\theta(X)-\lambda kl_\theta(X)],
    \qquad
    \mu_\theta(x)=\int \pi_\theta(t|x)m(t,x)d\nu(t).
\]
The benchmark rule and the candidate rule are fixed components of the estimand.
Thus the KL term depends on $P$ only through the marginal distribution of $X$.
The pathwise derivative of the $X$-marginal component is
\[
    \mathbb E[\{\mu_\theta(X)-\lambda kl_\theta(X)\}s_X(X)].
\]
The derivative of the conditional-mean component is
\[
    \mathbb E\left[
        \int \pi_\theta(t|X)
        \mathbb E[\{Y-m(t,X)\}s_Y(Y,T,X)|T=t,X]d\nu(t)
    \right].
\]
Using the observed-data treatment density, this term equals
\[
    \mathbb E\left[
        \frac{\pi_\theta(T|X)}{f(T|X)}\{Y-m(T,X)\}s_Y(Y,T,X)
    \right].
\]
There is no direct contribution from $s_p$, because the functional depends on
$P$ through $P_X$ and the conditional mean $m$, not through the treatment density itself.

Since $\mathbb E[s_X(X)]=0$, the first term can be centered as
\[
    \mathbb E[\{\mu_\theta(X)-\lambda kl_\theta(X)-W_\lambda(\theta)\}s_X(X)].
\]
Therefore
\[
    \frac{\partial}{\partial\varepsilon}
    W_{P_\varepsilon,\lambda}(\theta)\bigg|_{\varepsilon=0}
    =\mathbb E[\varphi_{\lambda,\theta}(Z)s(Z)],
\]
where
\[
    \varphi_{\lambda,\theta}(Z)
    =
    \frac{\pi_\theta(T|X)}{f(T|X)}\{Y-m(T,X)\}
    +\mu_\theta(X)-\lambda kl_\theta(X)-W_\lambda(\theta).
\]
This proves pathwise differentiability. The influence function is efficient:
in the nonparametric model the tangent space is the orthogonal sum of the three
score components, and $\varphi_{\lambda,\theta}$ lies in it. The residual
term is an outcome-regression direction, the centered term is a marginal
direction, and the treatment-density component is zero. Hence
$\varphi_{\lambda,\theta}$ is the projection of itself onto the tangent
space.

For the variance, the residual component has conditional mean zero given
$(T,X)$, while $\mu_\theta(X)-\lambda kl_\theta(X)-W_\lambda(\theta)$ is a
centered function of $X$. Hence the cross term is zero, and
\[
\begin{aligned}
    \Var\{\varphi_{\lambda,\theta}(Z)\}
    &=
    \mathbb E\left[
        \left\{
        \frac{\pi_\theta(T|X)}{f(T|X)}
        \right\}^2
        \{Y-m(T,X)\}^2
    \right]
    +
    \Var\{\mu_\theta(X)-\lambda kl_\theta(X)\}  \\
    &=
    \mathbb E\left[
        \frac{\pi_\theta(T|X)^2}{f(T|X)^2}
        \sigma^2(T,X)
    \right]
    +
    \Var\{\mu_\theta(X)-\lambda kl_\theta(X)\}.
\end{aligned}
\]
Finally, if $f(\cdot|\cdot)$ is known, the tangent space loses the
treatment-density component. Since $\varphi_{\lambda,\theta}$ has zero
projection on that component and the derivative representation above does not
involve $s_p$, the same function remains the efficient influence function, and
the bound is unchanged.
\end{proof}

\begin{proof}[Proof of Theorem 1]
Define
\[
    M(P,\theta)=\frac{\partial W_{P,\lambda}(\theta)}{\partial\theta}.
\]
Let $\{P_\varepsilon\}$ be any regular submodel with score $s(Z)$, and define
$\theta_\varepsilon=\theta^*(P_\varepsilon)$. Since $\theta_\varepsilon$ is an
interior maximizer of $W_{P_\varepsilon,\lambda}$ for all sufficiently small
$\varepsilon$,
\[
    M(P_\varepsilon,\theta_\varepsilon)=0,
    \qquad
    M(P,\theta^*)=0.
\]
Under Assumption 4 and the moment conditions of Lemma 1, the map
$(\varepsilon,\theta)\mapsto M(P_\varepsilon,\theta)$ is continuously
differentiable near $(0,\theta^*)$ and
$\partial M(P,\theta)/\partial\theta'|_{\theta=\theta^*}=-H$ is nonsingular.
By the implicit function theorem, $\varepsilon\mapsto\theta_\varepsilon$ is
well defined and differentiable for small $\varepsilon$. Differentiating the
first-order condition with respect to $\varepsilon$ at zero gives
\[
0=
\left.
\frac{\partial}{\partial\varepsilon}M(P_\varepsilon,\theta)
\right|_{\varepsilon=0,\theta=\theta^*}
+
\left.
\frac{\partial M(P,\theta)}{\partial\theta'}
\right|_{\theta=\theta^*}
\left.
\frac{\partial\theta_\varepsilon}{\partial\varepsilon}
\right|_{\varepsilon=0}.
\]
Since the second term uses
$\partial M(P,\theta)/\partial\theta'|_{\theta=\theta^*}=-H$, we obtain
\begin{align}
\label{eq:argmax-if}
\left.
\frac{\partial\theta_\varepsilon}{\partial\varepsilon}
\right|_{\varepsilon=0}
=H^{-1}
\left.
\frac{\partial}{\partial\varepsilon}M(P_\varepsilon,\theta)
\right|_{\varepsilon=0,\theta=\theta^*}.
\end{align}

By Lemma 1, for each fixed $\theta$,
\[
    \left.
    \frac{\partial}{\partial\varepsilon}W_{P_\varepsilon,\lambda}(\theta)
    \right|_{\varepsilon=0}
    =\mathbb E[\varphi_{\lambda,\theta}(Z)s(Z)].
\]
Under Assumption 4 and the moment conditions in Lemma 1, the map
$\theta\mapsto\varphi_{\lambda,\theta}$ is differentiable in $L_2(P)$ in a
neighborhood of $\theta^*$. Hence
\[
\left.
\frac{\partial}{\partial\varepsilon}M(P_\varepsilon,\theta)
\right|_{\varepsilon=0,\theta=\theta^*}
=\mathbb E\left[
\left.
\frac{\partial\varphi_{\lambda,\theta}(Z)}{\partial\theta}
\right|_{\theta=\theta^*}s(Z)
\right].
\]
Substituting this into \eqref{eq:argmax-if} gives the influence function
\[
    \mathrm{EIF}_{\theta^*}(Z)=
    H^{-1}
    \left.
    \frac{\partial\varphi_{\lambda,\theta}(Z)}{\partial\theta}
    \right|_{\theta=\theta^*}.
\]

Now
\[
\varphi_{\lambda,\theta}(Z)
=
\frac{\pi_\theta(T|X)}{f(T|X)}\{Y-m(T,X)\}
+\mu_\theta(X)-\lambda kl_\theta(X)-W_\lambda(\theta).
\]
Therefore
\[
\frac{\partial\varphi_{\lambda,\theta}(Z)}{\partial\theta}
=
\frac{Y-m(T,X)}{f(T|X)}\dot\pi_\theta(T|X)
+\dot\mu_\theta(X)-\lambda\dot{kl}_\theta(X)
-\frac{\partial W_\lambda(\theta)}{\partial\theta}.
\]
Evaluating at $\theta=\theta^*$ and using the first-order condition gives
\[
\left.
\frac{\partial\varphi_{\lambda,\theta}(Z)}{\partial\theta}
\right|_{\theta=\theta^*}
=
\frac{Y-m(T,X)}{f(T|X)}\dot\pi_{\theta^*}(T|X)
+\dot\mu_{\theta^*}(X)-\lambda\dot{kl}_{\theta^*}(X),
\]
where
\[
    \dot\mu_{\theta^*}(X)=\int m(t,X)\dot\pi_{\theta^*}(t|X)d\nu(t).
\]
This gives the displayed formula and the bound
\[
    V_{\mathrm{eff}}=H^{-1}\Var\left\{
    \left.
    \frac{\partial\varphi_{\lambda,\theta}(Z)}{\partial\theta}
    \right|_{\theta=\theta^*}
    \right\}H^{-1}.
\]

If $f(T|X)$ is known, the treatment-density tangent directions are
removed. The derivative above is orthogonal to those directions: the residual
term has conditional mean zero given $(T,X)$, and the other terms are functions
of $X$. Hence the same efficient influence function applies. The
H\'ajek--Le Cam convolution theorem
\citep[Theorem 25.20]{van1998asymptotic} then implies the stated convolution
representation for any regular estimator.

Finally, let $\hat\theta$ be regular and asymptotically linear with influence
function $\psi$, so that
\[
    \sqrt n(\hat\theta-\theta^*)=n^{-1/2}\sum_{i=1}^n\psi(Z_i)+o_p(1).
\]
Write $\psi=\mathrm{EIF}_{\theta^*}+\{\psi-\mathrm{EIF}_{\theta^*}\}$.
Regularity of an asymptotically linear estimator implies
\[
    \mathbb E[\psi(Z)s(Z)]
    =\mathbb E[\mathrm{EIF}_{\theta^*}(Z)s(Z)]
\]
for every score $s$ in the tangent space \citep[Section 25.3]{van1998asymptotic}.
Since the model is nonparametric, the tangent space is $L_0^2(P)$ and contains
$\mathrm{EIF}_{\theta^*}$. Therefore
$\psi-\mathrm{EIF}_{\theta^*}$ is orthogonal to
$\mathrm{EIF}_{\theta^*}$. The limit of $\sqrt n(\hat\theta-\theta^*)$ is
therefore the sum of two jointly Gaussian, uncorrelated, and hence independent
components: $G\sim N(0,V_{\mathrm{eff}})$ and a mean-zero Gaussian $U$ with
covariance $\Sigma_U=\Var\{\psi-\mathrm{EIF}_{\theta^*}\}$.
\end{proof}

\begin{proof}[Proof of Theorem 2]
Let $Z_n=\sqrt n(\hat\theta-\theta^*)$. A second-order Taylor expansion of
$W_\lambda$ around $\theta^*$ gives
\[
    W_\lambda(\theta^*)-W_\lambda(\hat\theta)
    =
    \frac12(\hat\theta-\theta^*)'H(\hat\theta-\theta^*)
    +o(\|\hat\theta-\theta^*\|^2).
\]
Hence
\[
    nR_\lambda(\hat\theta)=\frac12Z_n'HZ_n+o_p(1).
\]
From Theorem 1, $Z_n\Rightarrow Z=G+U$, with $G\sim N(0,V_{\mathrm{eff}})$ and
$U$ independent of $G$. Therefore
\[
    nR_\lambda(\hat\theta)\Rightarrow \frac12(G+U)'H(G+U).
\]
Let $\widetilde G=H^{1/2}G$ and $\widetilde U=H^{1/2}U$. Then
\[
\frac12(G+U)'H(G+U)=\frac12(\widetilde G+\widetilde U)'(\widetilde G+\widetilde U).
\]
The mean is
\[
    \frac12\tr\{H^{1/2}(V_{\mathrm{eff}}+\Sigma_U)H^{1/2}\}.
\]
For the variance, write the quadratic form as
\[
\frac12\widetilde G'\widetilde G+\widetilde G'\widetilde U+\frac12\widetilde U'\widetilde U.
\]
The covariance terms are zero by independence, mean zero, and the odd moments of
$\widetilde G$. Thus
\begin{align*}
\Var\left\{\frac12(G+U)'H(G+U)\right\}
&=\frac12\|H^{1/2}V_{\mathrm{eff}}H^{1/2}\|_F^2
+\tr(H^{1/2}V_{\mathrm{eff}}H\Sigma_UH^{1/2})
+\frac14\Var(U'HU)\\
&=\frac12\sum_{j=1}^p\ell_j^2
+\tr(H^{1/2}V_{\mathrm{eff}}H\Sigma_UH^{1/2})
+\frac14\Var(U'HU),
\end{align*}
where $\ell_j$ are the eigenvalues of $H^{1/2}V_{\mathrm{eff}}H^{1/2}$. If $U$ is Gaussian, then $G+U$ is Gaussian with covariance $V_{\mathrm{eff}}+\Sigma_U$, and the variance reduces to
\[
    \frac12\left\|H^{1/2}(V_{\mathrm{eff}}+\Sigma_U)H^{1/2}\right\|_F^2.
\]
\end{proof}

\begin{proof}[Proof of Theorem 3]
Let
\[
    \psi_{\lambda,\theta}^{tp}(Z)
    =
    \frac{\pi_\theta(T|X)Y}{f(T|X)}-\lambda kl_\theta(X),
    \qquad
    \widehat W_\lambda^{tp}(\theta)=\mathbb E_n\psi_{\lambda,\theta}^{tp}.
\]
Since $\mathbb E[\psi_{\lambda,\theta}^{tp}(Z)]=W_\lambda(\theta)$ and,
by Condition (S), the class
$\{\psi_{\lambda,\theta}^{tp}:\theta\in\Theta\}$ is Glivenko--Cantelli,
\[
    \sup_{\theta\in\Theta}|\widehat W_\lambda^{tp}(\theta)-W_\lambda(\theta)|\overset{p}{\to}0.
\]
By Assumption 3, the argmax theorem implies $\hat\theta^{tp}\overset{p}{\to}\theta^*$.

Let $\mathcal N$ be an open neighborhood of $\theta^*$ with
$\overline{\mathcal N}\subset\operatorname{int}(\Theta)$. With probability
approaching one, $\hat\theta^{tp}\in\mathcal N$ and the first-order condition
holds. A mean-value expansion gives
\begin{equation}
\label{eq:tp-score-expansion}
0=
\frac{\partial\widehat W_\lambda^{tp}(\theta^*)}{\partial\theta}
+
\frac{\partial^2\widehat W_\lambda^{tp}(\bar\theta_n)}{\partial\theta\partial\theta'}
(\hat\theta^{tp}-\theta^*),
\end{equation}
where $\bar\theta_n$ lies between $\hat\theta^{tp}$ and $\theta^*$. The Hessian
converges uniformly to $-H$ by Condition (S), and
\[
    \mathbb E\left[
    \left.
    \frac{\partial\psi_{\lambda,\theta}^{tp}(Z)}{\partial\theta}
    \right|_{\theta=\theta^*}
    \right]
    =
    \left.
    \frac{\partial W_\lambda(\theta)}{\partial\theta}
    \right|_{\theta=\theta^*}=0.
\]
Therefore
\[
\sqrt n(\hat\theta^{tp}-\theta^*)
=
H^{-1}(\mathbb E_n-\mathbb E)
\left[
    \left.
    \frac{\partial\psi_{\lambda,\theta}^{tp}(Z)}{\partial\theta}
    \right|_{\theta=\theta^*}
\right]
+o_p(1).
\]

To identify the inefficiency term, define
\[
    \eta^{tp}(Z)=
    \frac{m(T,X)}{f(T|X)}\dot\pi_{\theta^*}(T|X)-\dot\mu_{\theta^*}(X).
\]
Now
\[
    \left.
    \frac{\partial\psi_{\lambda,\theta}^{tp}(Z)}{\partial\theta}
    \right|_{\theta=\theta^*}
    =
    \frac{Y}{f(T|X)}\dot\pi_{\theta^*}(T|X)-\lambda\dot{kl}_{\theta^*}(X),
\]
whereas
\[
    \left.
    \frac{\partial\varphi_{\lambda,\theta}(Z)}{\partial\theta}
    \right|_{\theta=\theta^*}
    =
    \frac{Y-m(T,X)}{f(T|X)}\dot\pi_{\theta^*}(T|X)
    +\dot\mu_{\theta^*}(X)-\lambda\dot{kl}_{\theta^*}(X).
\]
Hence
\[
    \left.
    \frac{\partial\psi_{\lambda,\theta}^{tp}(Z)}{\partial\theta}
    \right|_{\theta=\theta^*}
    =
    \left.
    \frac{\partial\varphi_{\lambda,\theta}(Z)}{\partial\theta}
    \right|_{\theta=\theta^*}
    +\eta^{tp}(Z).
\]
The KL derivative cancels from the difference because it is common to the
true-propensity score and the efficient score.

Moreover, $\mathbb E[\eta^{tp}(Z)|X]=0$. Since the residual component of
$\partial\varphi_{\lambda,\theta}/\partial\theta$ has conditional mean zero
given $(T,X)$ and the remaining components are functions of $X$, we have
\[
    \mathbb E\left[
    \eta^{tp}(Z)
    \left(
    \left.
    \frac{\partial\varphi_{\lambda,\theta}(Z)}{\partial\theta}
    \right|_{\theta=\theta^*}
    \right)'
    \right]=0.
\]
Therefore
\[
\Var\left(
    \left.
    \frac{\partial\psi_{\lambda,\theta}^{tp}(Z)}{\partial\theta}
    \right|_{\theta=\theta^*}
\right)
=
\Var\left(
    \left.
    \frac{\partial\varphi_{\lambda,\theta}(Z)}{\partial\theta}
    \right|_{\theta=\theta^*}
\right)
+\Var\{\eta^{tp}(Z)\}.
\]
The summands are i.i.d. with mean zero and finite second moments under
Condition (S) and Assumptions 4--5. Hence the multivariate central limit
theorem applies to the stacked vector
\[
    \left(
    \left.
    \frac{\partial\varphi_{\lambda,\theta}(Z)}{\partial\theta}
    \right|_{\theta=\theta^*},
    \eta^{tp}(Z)
    \right),
\]
and the two components of the limit are jointly Gaussian. It follows that
\[
    \sqrt n(\hat\theta^{tp}-\theta^*)\Rightarrow G+U^{tp},
\]
where $G\sim N(0,V_{\mathrm{eff}})$ and
\[
    \Sigma_U^{tp}=H^{-1}\Var\{\eta^{tp}(Z)\}H^{-1}.
\]
The two components are independent because their joint limit is Gaussian and
their covariance is zero. The regret result follows by applying Theorem 2 with
$\Sigma_U=\Sigma_U^{tp}$.
\end{proof}

\begin{proof}[Proof of Lemma 2]
Only parts (1)--(4) of Assumption 6 are used in this proof. After a nonsingular linear transformation of each basis, which the eigenvalue
bounds in Assumption 6(3) keep uniformly well conditioned and which leaves the
estimated weights unchanged, we may normalize
\[
    \int u_{K_1}(t)u_{K_1}(t)'d\nu(t)=I_{K_1},
    \qquad
    \mathbb E[v_{K_2}(X)v_{K_2}(X)']=I_{K_2}.
\]
Write
\[
    \gamma=(\rho')^{-1}(\omega),
    \qquad
    b_\Lambda(t,x)=u_{K_1}(t)'\Lambda v_{K_2}(x),
\]
and
\[
    G_K(\Lambda)
    =\mathbb E[\rho\{b_\Lambda(T,X)\}]
    -\left(\int u_{K_1}(t)d\nu(t)\right)'\Lambda\mathbb E[v_{K_2}(X)].
\]
Let $\widehat G_K$ denote the sample analogue, let $\Lambda_K^*$ and
$\widehat\Lambda_K$ be the population and sample maximizers, and define
\[
    \omega_K^*(t,x)=\rho'\{b_{\Lambda_K^*}(t,x)\},
    \qquad
    \widehat\omega_K(t,x)=\rho'\{b_{\widehat\Lambda_K}(t,x)\}.
\]
Since $\rho(v)=-e^{-v-1}$, $\rho'=-\rho''=\rho'''>0$. Also
$1/\overline f\le\omega\le1/\underline f$ by Assumption 6(1), so
$\gamma$ lies in a fixed compact interval. Let $\Gamma$ be a one-unit
enlargement of that interval, and let
\[
    \overline a=\sup_{\Gamma}\rho',
    \qquad
    \underline a=\inf_{\Gamma}\rho'>0.
\]

We use two identities repeatedly. First, the balancing identity for the true
weight gives, for every matrix $A$,
\begin{align}
    \mathbb E\left[\omega(T,X)
    \{u_{K_1}(T)'Av_{K_2}(X)\}^2\right]
    =\|A\|_F^2 .
    \label{eq:gram-identity}
\end{align}
Second, since $\underline f\le f\le\overline f$,
\begin{align}
    \underline f\|A\|_F^2
    \le
    \mathbb E[\{u_{K_1}(T)'Av_{K_2}(X)\}^2]
    \le
    \overline f\|A\|_F^2 .
    \label{eq:norm-equivalence}
\end{align}
Also, uniformly in $(t,x)$,
\[
    |u_{K_1}(t)'Av_{K_2}(x)|\le\zeta(K)\|A\|_F.
\]

By Assumption 6(2), there is a matrix $\Lambda_K$ such that
$\|\gamma-b_{\Lambda_K}\|_\infty\le CK^{-\alpha}$. Thus
$b_{\Lambda_K}$ takes values in $\Gamma$ for large $K$, and
\[
    \sup_{t,x}|\omega(t,x)-\rho'\{b_{\Lambda_K}(t,x)\}|
    \le C K^{-\alpha}.
\]
The balancing identity gives
\[
    \nabla G_K(\Lambda_K)
    =\mathbb E[\{\rho'(b_{\Lambda_K})-\omega\}
        u_{K_1}(T)v_{K_2}(X)'].
\]
For any $A$ with $\|A\|_F=1$, Cauchy--Schwarz and
\eqref{eq:gram-identity} imply
\[
\begin{aligned}
    |\langle\nabla G_K(\Lambda_K),A\rangle|
    &\le
    \left(
    \mathbb E[\{\rho'(b_{\Lambda_K})-\omega\}^2 f(T|X)]
    \right)^{1/2}
    \left(
    \mathbb E[\omega\{u_{K_1}(T)'Av_{K_2}(X)\}^2]
    \right)^{1/2}  \\
    &\le C K^{-\alpha}.
\end{aligned}
\]
Hence $\|\nabla G_K(\Lambda_K)\|_F\le CK^{-\alpha}$.

Consider the ball $\|\Lambda-\Lambda_K\|_F\le C_1K^{-\alpha}$. On this ball,
\[
    \sup_{t,x}|b_\Lambda(t,x)-b_{\Lambda_K}(t,x)|
    \le C_1\zeta(K)K^{-\alpha}=o(1),
\]
because Assumptions 6(2) and 6(4) imply $\zeta(K)K^{-\alpha}\to0$. Thus
$b_\Lambda$ remains in $\Gamma$ for large $K$. For every $\Delta$,
\[
    D^2G_K(\Lambda)[\Delta,\Delta]
    =\mathbb E[\rho''\{b_\Lambda(T,X)\}
    \{u_{K_1}(T)'\Delta v_{K_2}(X)\}^2]
    \le-\underline a\underline f\|\Delta\|_F^2,
\]
where the last step uses \eqref{eq:norm-equivalence}. A second-order expansion
around $\Lambda_K$ shows that, for $C_1$ large enough,
\[
    G_K(\Lambda)<G_K(\Lambda_K)
    \qquad\text{on } \|\Lambda-\Lambda_K\|_F=C_1K^{-\alpha}.
\]
Concavity therefore places the maximizer inside this ball. Strict concavity on
this neighborhood gives uniqueness, and
\[
    \|\Lambda_K^*-\Lambda_K\|_F=O(K^{-\alpha}).
\]
By the Lipschitz property of $\rho'$ on $\Gamma$ and
\eqref{eq:norm-equivalence},
\[
    \int |\omega_K^*(t,x)-\omega(t,x)|^2dF_{T,X}(t,x)=O(K^{-2\alpha}),
    \qquad
    \mathbb E_n|\omega_K^*(T,X)-\omega(T,X)|^2=O_p(K^{-2\alpha}).
\]

We next control the sample maximizer. Define
\[
    \widehat Q_n=\mathbb E_n\left[\omega(T,X)
    \{v_{K_2}(X)v_{K_2}(X)'\}\otimes
    \{u_{K_1}(T)u_{K_1}(T)'\}\right].
\]
Then $\mathbb E\widehat Q_n=I_K$ by \eqref{eq:gram-identity}. Also,
using $\omega^2\le\omega/\underline f$,
\[
    \mathbb E\|\widehat Q_n-I_K\|_F^2
    \le
    \frac{1}{n}\mathbb E[\omega^2\|u_{K_1}(T)\|^4\|v_{K_2}(X)\|^4]
    \le
    \frac{\zeta(K)^2K}{\underline f n}\to0 .
\]
Thus, on an event whose probability tends to one,
\begin{align}
    \frac12\|A\|_F^2
    \le
    \mathbb E_n[\omega(T,X)\{u_{K_1}(T)'Av_{K_2}(X)\}^2]
    \le
    \frac32\|A\|_F^2
    \qquad\text{for every }A .
    \label{eq:empirical-equivalence}
\end{align}

The score $\nabla\widehat G_K(\Lambda_K^*)$ has mean zero, and
\[
    \mathbb E\|\nabla\widehat G_K(\Lambda_K^*)\|_F^2
    \le C K/n.
\]
Indeed, the first part of the score is bounded using
$\rho'(b_{\Lambda_K^*})\le\overline a$ and
$\mathbb E[\|u_{K_1}(T)\|^2\|v_{K_2}(X)\|^2]\le\overline f K$; the second
part is bounded by
$\|\int u_{K_1}d\nu\|^2\mathbb E\|v_{K_2}(X)\|^2\le
\nu(\mathcal T)K$.
Now consider the ball $\|\Lambda-\Lambda_K^*\|_F\le C_2\sqrt{K/n}$. On this
ball,
\[
    \sup_{t,x}|b_\Lambda(t,x)-b_{\Lambda_K^*}(t,x)|
    \le C_2\zeta(K)\sqrt{K/n}=o(1)
\]
by Assumption 6(4), so $b_\Lambda$ remains in $\Gamma$ for large $n$. On the
event \eqref{eq:empirical-equivalence}, the sample Hessian satisfies
\[
    D^2\widehat G_K(\Lambda)[\Delta,\Delta]
    \le-\frac{\underline a\underline f}{2}\|\Delta\|_F^2.
\]
Chebyshev's inequality and the same boundary comparison as above then imply
\[
    \|\widehat\Lambda_K-\Lambda_K^*\|_F=O_p(\sqrt{K/n}).
\]
Moreover, with probability approaching one, the sample maximizer is unique,
interior, and satisfies the first-order conditions. Equivalently, the exact
in-sample balancing equations hold on this event.

Finally, on the same event, the Lipschitz property of $\rho'$ on $\Gamma$ and
\eqref{eq:norm-equivalence}--\eqref{eq:empirical-equivalence} give
\[
    \int|\widehat\omega_K-\omega_K^*|^2dF_{T,X}=O_p(K/n),
    \qquad
    \mathbb E_n|\widehat\omega_K-\omega_K^*|^2=O_p(K/n).
\]
Combining these bounds with the population approximation, and using
$\sqrt nK^{-\alpha}\to0$, gives both displays of the lemma.
\end{proof}

\begin{proof}[Proof of Theorem 4]
Write the outcome-welfare part of the estimated-propensity criterion as
\[
    \widehat W^{ep}_0(\theta)
    =
    \mathbb E_n[\widehat\omega_K(T,X)\pi_\theta(T|X)Y],
    \qquad
    \widehat W^{ep}_\lambda(\theta)
    =\widehat W^{ep}_0(\theta)-\lambda\mathbb E_n kl_\theta(X).
\]

We first show consistency. By Lemma 2,
\[
    \mathbb E_n[\{\widehat\omega_K(T,X)-\omega(T,X)\}^2]=O_p(K/n)=o_p(1).
\]
Together with Condition (S), bounded outcomes, and the uniform law of large
numbers for the smooth finite-dimensional class, this gives
\[
    \sup_{\theta\in\Theta}
    |\widehat W^{ep}_0(\theta)-W_0(\theta)|=o_p(1).
\]
The KL term is a known smooth function of $(X,\theta)$, so Condition (S) also
gives
\[
    \sup_{\theta\in\Theta}
    |\mathbb E_n kl_\theta(X)-\mathbb E kl_\theta(X)|=o_p(1).
\]
Therefore
\[
    \sup_{\theta\in\Theta}
    |\widehat W^{ep}_\lambda(\theta)-W_\lambda(\theta)|=o_p(1),
\]
and Assumption 3 gives $\widehat\theta^{ep}\overset{p}{\to}\theta^*$.

It remains to derive the local score expansion. Let
\[
    \dot\pi_*(t|x)
    =
    \left.
    \frac{\partial\pi_\theta(t|x)}{\partial\theta}
    \right|_{\theta=\theta^*},
    \qquad
    \dot\mu_*(x)=\int m(t,x)\dot\pi_*(t|x)d\nu(t).
\]
The outcome-welfare score at $\theta^*$ is
\[
    \left.
    \frac{\partial\widehat W^{ep}_0(\theta)}{\partial\theta}
    \right|_{\theta=\theta^*}
    =\mathbb E_n[\widehat\omega_K(T,X)\dot\pi_*(T|X)Y].
\]
Subtract the efficient outcome-welfare score
\[
    \mathbb E_n[
        \omega(T,X)\dot\pi_*(T|X)\{Y-m(T,X)\}
        +\dot\mu_*(X)
    ].
\]
The difference is $A_n+B_n$, where
\[
    A_n=\mathbb E_n[
        \{\widehat\omega_K(T,X)-\omega(T,X)\}
        \dot\pi_*(T|X)\{Y-m(T,X)\}],
\]
and
\[
    B_n=
    \mathbb E_n[\widehat\omega_K(T,X)m(T,X)\dot\pi_*(T|X)]
    -\mathbb E_n[\dot\mu_*(X)].
\]
For $A_n$, condition on the treatment and covariates. The estimated weights
are functions only of $(T_i,X_i)_{i=1}^n$, while
$\mathbb E[Y-m(T,X)|T,X]=0$. Therefore
\[
    \mathbb E[n\|A_n\|^2|(T_i,X_i)_{i=1}^n]
    \lesssim
    \mathbb E_n[\{\widehat\omega_K(T,X)-\omega(T,X)\}^2]
    =O_p(K/n),
\]
and hence $\sqrt nA_n=o_p(1)$.

We next control $B_n$ componentwise. For $j=1,\ldots,p$, write
\[
    \dot\pi_{*,j}(t|x)
    =
    \left.
    \frac{\partial\pi_\theta(t|x)}{\partial\theta_j}
    \right|_{\theta=\theta^*}.
\]
By Assumption 6(5), choose
$\Lambda_{j,K}\in\mathbb R^{K_1\times K_2}$ such that, with
\[
    r_{j,K}(t,x)
    =
    m(t,x)\dot\pi_{*,j}(t|x)
    -u_{K_1}(t)'\Lambda_{j,K}v_{K_2}(x),
\]
we have
\[
    \max_{1\le j\le p}
    \|r_{j,K}\|_{L_2(\nu\otimes P_X)}=o(K^{-1/2}).
\]
On the event from the proof of Lemma 2 whose probability tends to one, the
exact in-sample balancing equations imply
\[
    \mathbb E_n[\widehat\omega_K(T,X)
    u_{K_1}(T)'\Lambda_{j,K}v_{K_2}(X)]
    =
    \mathbb E_n\left[\int
    u_{K_1}(t)'\Lambda_{j,K}v_{K_2}(X)d\nu(t)\right].
\]
Therefore the $j$th component of $B_n$ satisfies
\[
\begin{aligned}
    B_{n,j}
    &=
    \mathbb E_n[
        \{\widehat\omega_K(T,X)-\omega(T,X)\}r_{j,K}(T,X)]  \\
    &\quad+
    \mathbb E_n\left[
        \omega(T,X)r_{j,K}(T,X)
        -\int r_{j,K}(t,X)d\nu(t)
    \right].
\end{aligned}
\]
Since
\[
    \mathbb E[r_{j,K}(T,X)^2]
    =\mathbb E_X\int r_{j,K}(t,X)^2f(t|X)d\nu(t)
    \le\overline f\|r_{j,K}\|_{L_2(\nu\otimes P_X)}^2,
\]
Markov's inequality gives
$\mathbb E_n r_{j,K}(T,X)^2=O_p(\|r_{j,K}\|_{L_2(\nu\otimes P_X)}^2)$.
By Cauchy--Schwarz and Lemma 2, the first term is, after multiplication by
$\sqrt n$,
\[
    O_p\{\sqrt K\,
    \|r_{j,K}\|_{L_2(\nu\otimes P_X)}\}=o_p(1).
\]
The second term has mean zero because
\[
    \mathbb E[\omega(T,X)r_{j,K}(T,X)|X]
    =\int r_{j,K}(t,X)d\nu(t),
\]
and its variance is bounded by a constant times
$\|r_{j,K}\|_{L_2(\nu\otimes P_X)}^2$. Hence its $\sqrt n$-scaled empirical
average is also $o_p(1)$. Since the dimension of $\theta$ is fixed, $\sqrt nB_n=o_p(1)$.
Thus
\[
    \sqrt n
    \left.
    \frac{\partial\widehat W^{ep}_0(\theta)}{\partial\theta}
    \right|_{\theta=\theta^*}
    =
    \frac1{\sqrt n}\sum_{i=1}^n
    \left\{
        \omega(T_i,X_i)\dot\pi_*(T_i|X_i)
        \{Y_i-m(T_i,X_i)\}
        +\dot\mu_*(X_i)
    \right\}
    +o_p(1).
\]

The penalized score adds $-\lambda\mathbb E_n\dot{kl}_{\theta^*}(X)$. Since
\[
    \left.
    \frac{\partial W_\lambda(\theta)}{\partial\theta}
    \right|_{\theta=\theta^*}
    =\mathbb E[\dot\mu_*(X)-\lambda\dot{kl}_{\theta^*}(X)]=0,
\]
we obtain
\[
    \sqrt n
    \left.
    \frac{\partial\widehat W^{ep}_\lambda(\theta)}{\partial\theta}
    \right|_{\theta=\theta^*}
    =
    \frac1{\sqrt n}\sum_{i=1}^n
    \left.
    \frac{\partial\varphi_{\lambda,\theta}(Z_i)}{\partial\theta}
    \right|_{\theta=\theta^*}
    +o_p(1).
\]

For the Hessian, write, for $\theta$ in a neighborhood
$\overline{\mathcal N}$ of $\theta^*$,
\[
\begin{aligned}
    \frac{\partial^2\widehat W^{ep}_\lambda(\theta)}{\partial\theta\partial\theta'}
    &=
    \mathbb E_n\left[
        \omega(T,X)
        \frac{\partial^2\pi_\theta(T|X)}{\partial\theta\partial\theta'}Y
    \right] \\
    &\quad+
    \mathbb E_n\left[
        \{\widehat\omega_K(T,X)-\omega(T,X)\}
        \frac{\partial^2\pi_\theta(T|X)}{\partial\theta\partial\theta'}Y
    \right]
    -\lambda\mathbb E_n
    \frac{\partial^2kl_\theta(X)}{\partial\theta\partial\theta'}.
\end{aligned}
\]
By Cauchy--Schwarz, Lemma 2, and Condition (S), the middle term is bounded in
norm, uniformly over $\overline{\mathcal N}$, by
\[
    \{\mathbb E_n(\widehat\omega_K-\omega)^2\}^{1/2}
    \left\{
    \mathbb E_n\sup_{\theta\in\overline{\mathcal N}}
    \left\|
        \frac{\partial^2\pi_\theta(T|X)}{\partial\theta\partial\theta'}Y
    \right\|^2
    \right\}^{1/2}
    =O_p(\sqrt{K/n})=o_p(1).
\]
The first and third terms converge uniformly on $\overline{\mathcal N}$ to the
corresponding population Hessians by Condition (S), and the population Hessian
is continuous at $\theta^*$ with value $-H$ by Assumption 3. Hence
\[
    \left.
    \frac{\partial^2\widehat W^{ep}_\lambda(\theta)}{\partial\theta\partial\theta'}
    \right|_{\theta=\bar\theta_n}
    =-H+o_p(1)
\]
for every $\bar\theta_n\overset{p}{\to}\theta^*$.

The first-order condition and a mean-value expansion now imply
\[
    \sqrt n(\widehat\theta^{ep}-\theta^*)
    =
    H^{-1}
    \frac1{\sqrt n}\sum_{i=1}^n
    \left.
    \frac{\partial\varphi_{\lambda,\theta}(Z_i)}{\partial\theta}
    \right|_{\theta=\theta^*}
    +o_p(1).
\]
Therefore $\sqrt n(\widehat\theta^{ep}-\theta^*)\Rightarrow G$, where
$G\sim N(0,V_{\mathrm{eff}})$. A second-order expansion of $W_\lambda$ around
$\theta^*$ gives
\[
    nR_\lambda(\widehat\theta^{ep})\Rightarrow \frac12G'HG.
\]
This proves the theorem.
\end{proof}

\begin{proof}[Proof of Theorem 5]
Write the outcome-welfare part of the cross-fitted doubly robust criterion as
\[
\widehat W^{dr}_0(\theta)
=
\frac1n\sum_{\ell=1}^L\sum_{i\in\mathcal I_\ell}
\left[
    \widehat\omega^{(-\ell)}(T_i,X_i)\pi_\theta(T_i|X_i)
    \{Y_i-\widehat m^{(-\ell)}(T_i,X_i)\}
    +
    \int \widehat m^{(-\ell)}(t,X_i)\pi_\theta(t|X_i)d\nu(t)
\right].
\]
The penalized criterion is
\[
    \widehat W^{dr}_\lambda(\theta)
    =\widehat W^{dr}_0(\theta)-\lambda\mathbb E_n kl_\theta(X).
\]

We first note consistency. Fix a fold $\ell$ and condition on the training
sample $\mathcal I_\ell^c$. For any fixed $\theta$, the conditional bias of
the outcome-welfare part is
\[
\begin{aligned}
&\mathbb E[
    \widehat\omega^{(-\ell)}(T,X)\pi_\theta(T|X)
    \{Y-\widehat m^{(-\ell)}(T,X)\}
    +
    \widehat\mu_\theta^{(-\ell)}(X)
    |\mathcal I_\ell^c]
    -W_0(\theta)                                      \\
&\quad =
\mathbb E[
    \{\widehat\omega^{(-\ell)}(T,X)-\omega(T,X)\}
    \pi_\theta(T|X)
    \{m(T,X)-\widehat m^{(-\ell)}(T,X)\}
    |\mathcal I_\ell^c].
\end{aligned}
\]
The two first-order bias terms vanish because
$\mathbb E[Y-m(T,X)|T,X]=0$ and
$\mathbb E[\omega(T,X)h(T,X)|X]=\int h(t,X)d\nu(t)$. By
Cauchy--Schwarz, the remaining bias is bounded uniformly in $\theta$ by a
constant times
\[
    \|\widehat\omega^{(-\ell)}-\omega\|_{L_2}
    \|\widehat m^{(-\ell)}-m\|_{L_2}=o_p(n^{-1/2}).
\]
Conditionally on $\mathcal I_\ell^c$, the class indexed by $\theta$ has a
bounded envelope and is continuous in $\theta$ by Condition (S) and the bounded
versions of the nuisance estimates in Assumption 7(2). Hence the centered
empirical process is $o_p(1)$ uniformly over $\Theta$ fold by fold, and the KL
term satisfies a uniform law of large numbers. Therefore
\[
    \sup_{\theta\in\Theta}
    |\widehat W^{dr}_\lambda(\theta)-W_\lambda(\theta)|=o_p(1),
\]
and Assumption 3 gives $\widehat\theta^{dr}\overset{p}{\to}\theta^*$.

We now derive the score expansion. Let
\[
    \dot\pi_*(t|x)=\left.
    \frac{\partial\pi_\theta(t|x)}{\partial\theta}
    \right|_{\theta=\theta^*},
    \qquad
    \dot\mu_*(x)=\int m(t,x)\dot\pi_*(t|x)d\nu(t).
\]
For a fold $\ell$, condition again on $\mathcal I_\ell^c$. The derivative of
the outcome-welfare part at $\theta^*$, minus the efficient outcome-welfare
score, is the sum of
\[
\begin{aligned}
A_{\ell n}
&=
\mathbb E_{n,\ell}[
    \{\widehat\omega^{(-\ell)}-\omega\}(T,X)
    \dot\pi_*(T|X)\{Y-m(T,X)\}],\\
B_{\ell n}
&=
\mathbb E_{n,\ell}\left[
    \int \{\widehat m^{(-\ell)}(t,X)-m(t,X)\}
    \dot\pi_*(t|X)d\nu(t)\right.\\
&\hspace{3.0cm}\left.
    -\omega(T,X)\dot\pi_*(T|X)
    \{\widehat m^{(-\ell)}(T,X)-m(T,X)\}
\right],\\
C_{\ell n}
&=
\mathbb E_{n,\ell}[
    \{\widehat\omega^{(-\ell)}-\omega\}(T,X)
    \dot\pi_*(T|X)
    \{m(T,X)-\widehat m^{(-\ell)}(T,X)\}],
\end{aligned}
\]
where $\mathbb E_{n,\ell}$ denotes the average over observations in fold
$\ell$.

The term $\sqrt nA_{\ell n}$ is conditionally mean zero with conditional
variance bounded by a constant times
$\|\widehat\omega^{(-\ell)}-\omega\|_{L_2}^2=o_p(1)$, so
$\sqrt nA_{\ell n}=o_p(1)$ by the conditional Chebyshev inequality. The term
$\sqrt nB_{\ell n}$ is also conditionally mean zero. Because
$\dot\pi_{\theta^*}$ is supported on the benchmark support, where
$f\ge\underline f$ by Assumption 5, and is bounded with bounded
$\nu$-integral by Condition (S), its conditional variance is bounded by a
constant times $\|\widehat m^{(-\ell)}-m\|_{L_2}^2=o_p(1)$, and
$\sqrt nB_{\ell n}=o_p(1)$ by the same argument. Finally, the conditional
mean of $C_{\ell n}$ is bounded by Cauchy--Schwarz by a constant times
\[
    \|\widehat\omega^{(-\ell)}-\omega\|_{L_2}
    \|\widehat m^{(-\ell)}-m\|_{L_2}=o_p(n^{-1/2}).
\]
Using the bounded versions in Assumption 7(2), the centered part of
$\sqrt nC_{\ell n}$ has conditional variance bounded by a constant times
$\|\widehat\omega^{(-\ell)}-\omega\|_{L_2}^2=o_p(1)$. Hence
$\sqrt nC_{\ell n}=o_p(1)$.

Since the number of folds is fixed, summing over folds gives
\[
    \sqrt n
    \left.
    \frac{\partial\widehat W^{dr}_0(\theta)}{\partial\theta}
    \right|_{\theta=\theta^*}
    =
    \frac1{\sqrt n}\sum_{i=1}^n
    \left\{
        \omega(T_i,X_i)\dot\pi_*(T_i|X_i)
        \{Y_i-m(T_i,X_i)\}
        +\dot\mu_*(X_i)
    \right\}
    +o_p(1).
\]
Adding the penalty score and using the first-order condition
\[
    \mathbb E[\dot\mu_*(X)-\lambda\dot{kl}_{\theta^*}(X)]=0
\]
yields
\[
    \sqrt n
    \left.
    \frac{\partial\widehat W^{dr}_\lambda(\theta)}{\partial\theta}
    \right|_{\theta=\theta^*}
    =
    \frac1{\sqrt n}\sum_{i=1}^n
    \left.
    \frac{\partial\varphi_{\lambda,\theta}(Z_i)}{\partial\theta}
    \right|_{\theta=\theta^*}
    +o_p(1).
\]
The same orthogonality and product-rate argument applied in a shrinking
neighborhood of $\theta^*$, together with Condition (S) and the smooth
sample-average KL term, gives
\[
    \left.
    \frac{\partial^2\widehat W^{dr}_\lambda(\theta)}{\partial\theta\partial\theta'}
    \right|_{\theta=\bar\theta_n}
    =-H+o_p(1),
    \qquad
    \bar\theta_n\overset{p}{\to}\theta^*.
\]
The first-order condition and a mean-value expansion therefore imply
\[
    \sqrt n(\widehat\theta^{dr}-\theta^*)
    =
    H^{-1}
    \frac1{\sqrt n}\sum_{i=1}^n
    \left.
    \frac{\partial\varphi_{\lambda,\theta}(Z_i)}{\partial\theta}
    \right|_{\theta=\theta^*}
    +o_p(1).
\]
Thus $\sqrt n(\widehat\theta^{dr}-\theta^*)\Rightarrow G$, where
$G\sim N(0,V_{\mathrm{eff}})$. The second-order expansion of $W_\lambda$
around $\theta^*$ gives
\[
    nR_\lambda(\widehat\theta^{dr})\Rightarrow \frac12G'HG.
\]
This completes the proof.
\end{proof}

\begin{proof}[Proof of Theorem 6]
Throughout the proof, $\mathbb E_n$ denotes the sample average and
$\mathbb E$ the population expectation. Constants may change from line to
line. We take $\mathbf\Pi$ to be pointwise measurable. Under Assumption 8,
$p(x)\in[\kappa,1-\kappa]$, and hence the inverse propensity weights,
$Y$, $m_1$, and $m_0$ are uniformly bounded.

For reference, write the two influence functions as
\[
\begin{aligned}
    \varphi^{tp}_\pi(Z)
    &=
    \left[
        \omega_1(X)TY
        -
        \omega_0(X)(1-T)Y
    \right]\pi(X)
    +
    \omega_0(X)(1-T)Y
    -
    W(\pi),\\
    \varphi^{ep}_\pi(Z)
    &=
    \left[
        \omega_1(X)T\{Y-m_1(X)\}
        -
        \omega_0(X)(1-T)\{Y-m_0(X)\}
        +m_1(X)-m_0(X)
    \right]\pi(X)\\
    &\quad
    +
    \omega_0(X)(1-T)\{Y-m_0(X)\}
    +m_0(X)-W(\pi).
\end{aligned}
\]

First consider the true-propensity estimator. By construction,
\[
    \sqrt n\{\widehat W^{tp}(\pi)-W(\pi)\}
    =\frac1{\sqrt n}\sum_{i=1}^n\varphi^{tp}_\pi(Z_i),
    \qquad \pi\in\mathbf\Pi .
\]
The class $\{\varphi^{tp}_\pi:\pi\in\mathbf\Pi\}$ is a bounded VC-type class, because it is obtained from $\mathbf\Pi$ by multiplying by bounded fixed functions and adding a bounded constant term. Therefore it is $P$-Donsker by Lemma~\ref{lm:covering-number-VC} and the standard uniform-entropy Donsker theorem. The same argument applies to $\{\varphi^{ep}_\pi:\pi\in\mathbf\Pi\}$. Hence
\[
    \sqrt n\{\widehat W^{tp}(\cdot)-W(\cdot)\}
    \Rightarrow G_{tp}
    \quad\text{in }\ell^\infty(\mathbf\Pi),
\]
where the covariance function of $G_{tp}$ is induced by
$\varphi^{tp}_\pi$. The first display of the theorem follows from the
continuous mapping theorem. The same Donsker statement will be used below for
$\varphi^{ep}_\pi$.

We next turn to the estimated-propensity estimator. With binary treatment the
balancing program separates by treatment arm. On the event from Lemma 2, whose
probability tends to one, the dual solution is interior and the first-order
conditions imply
\begin{align}
\label{eqn:balance}
    \mathbb E_n[T\widehat\omega_1v]=\mathbb E_n[v],
    \qquad
    \mathbb E_n[(1-T)\widehat\omega_0v]=\mathbb E_n[v],
    \qquad v\in V_K .
\end{align}
Lemma 2 also gives
\[
    \mathbb E_n[T(\widehat\omega_1-\omega_1)^2
    +(1-T)(\widehat\omega_0-\omega_0)^2]=O_p(K/n).
\]

For $\pi\in\mathbf\Pi$, write
\[
    g_\pi(x)=m_1(x)\pi(x),
    \qquad
    h_\pi(x)=m_0(x)\{1-\pi(x)\}.
\]
Let $g_{K,\pi}$ and $h_{K,\pi}$ be their $L_2(P_X)$ projections on
$V_K$, and define the residuals
$r_{1K,\pi}=g_\pi-g_{K,\pi}$ and
$r_{0K,\pi}=h_\pi-h_{K,\pi}$. Assumption 9 gives
\[
    \sup_{\pi\in\mathbf\Pi}
    \max\{\|r_{1K,\pi}\|_{L_2(P_X)},
          \|r_{0K,\pi}\|_{L_2(P_X)}\}
    =a_K,
    \qquad
    a_K\sqrt{K\log n}\to0,
\]
and $\zeta(K)K\log n/\sqrt n\to0$. The residual classes are still VC-type up
to the $K$-dimensional sieve projection. In particular, Lemma~\ref{lm:covering-number-VC} and the standard maximal
inequality for VC-type classes gives, for $j=0,1$,
\begin{equation}\label{eqn:residual-maximal}
\begin{aligned}
    \sup_{\pi\in\mathbf\Pi}
    \left|\sqrt n(\mathbb E_n-\mathbb E)
    \{(T\omega_1-1)r_{1K,\pi}\}\right|
    &=o_p(1),
    \\
    \sup_{\pi\in\mathbf\Pi}
    \mathbb E_n r_{jK,\pi}^2
    &=O_p\left(a_K^2+\frac{\zeta(K)^2K\log n}{n}\right).
\end{aligned}
\end{equation}
The same bound holds with $T\omega_1-1$ replaced by
$(1-T)\omega_0-1$.

A direct decomposition gives
\[
    \sqrt n\{\widehat W^{ep}(\pi)-W(\pi)\}
    -\sqrt n\mathbb E_n\varphi^{ep}_\pi
    =A_{n,\pi}+B_{n,\pi},
\]
where
\[
\begin{aligned}
    A_{n,\pi}
    &=\sqrt n\mathbb E_n\big[(T\widehat\omega_1-1)g_\pi
    +\{(1-T)\widehat\omega_0-1\}h_\pi\big],\\
    B_{n,\pi}
    &=\sqrt n\mathbb E_n\big[(\widehat\omega_1-\omega_1)T(Y-m_1)\pi
    +(\widehat\omega_0-\omega_0)(1-T)(Y-m_0)(1-\pi)\big].
\end{aligned}
\]
We only show the treated-arm bounds; the control arm is identical.

By the balance equations in \eqref{eqn:balance},
\[
\begin{aligned}
    \sqrt n\mathbb E_n[(T\widehat\omega_1-1)g_\pi]
    &=\sqrt n\mathbb E_n[(T\widehat\omega_1-1)r_{1K,\pi}]\\
    &=\sqrt n\mathbb E_n[(T\omega_1-1)r_{1K,\pi}]
      +\sqrt n\mathbb E_n[T(\widehat\omega_1-\omega_1)r_{1K,\pi}].
\end{aligned}
\]
The first term is $o_p(1)$ uniformly in $\pi$ by
\eqref{eqn:residual-maximal}. For the second term, Cauchy--Schwarz, Lemma 2,
and \eqref{eqn:residual-maximal} give
\[
\begin{aligned}
    &\sup_{\pi\in\mathbf\Pi}
    \left|\sqrt n\mathbb E_n[T(\widehat\omega_1-\omega_1)r_{1K,\pi}]\right|\le
    O_p(\sqrt K) \sup_{\pi\in\mathbf\Pi}\{\mathbb E_n r_{1K,\pi}^2\}^{1/2}
    =o_p(1).
\end{aligned}
\]
Thus $\sup_{\pi\in\mathbf\Pi}|A_{n,\pi}|=o_p(1)$.

For $B_{n,\pi}$, condition on the treatment and covariates. The estimated
weights are functions only of $(T_i,X_i)_{i=1}^n$, and the residuals
$Y_i-m_t(X_i)$ have conditional mean zero. By the usual symmetrization and
VC entropy bound, or equivalently by Dudley's inequality with the covering bound of Lemma~\ref{lm:covering-number-VC}, conditionally on
$(T_i,X_i)_{i=1}^n$,
\[
\begin{aligned}
    \mathbb E\left[
    \sup_{\pi\in\mathbf\Pi}
    \left|\sqrt n\mathbb E_n[(\widehat\omega_1-\omega_1)T(Y-m_1)\pi]
    \right|
    \Bigm| (T_i,X_i)_{i=1}^n\right]
    \le
    C\{\mathbb E_nT(\widehat\omega_1-\omega_1)^2\}^{1/2}.
\end{aligned}
\]
The right-hand side is $O_p(\sqrt{K/n})=o_p(1)$. Markov's inequality and the
control-arm analogue imply
$\sup_{\pi\in\mathbf\Pi}|B_{n,\pi}|=o_p(1)$. Therefore
\[
    \sup_{\pi\in\mathbf\Pi}
    \left|
    \sqrt n\{\widehat W^{ep}(\pi)-W(\pi)\}
    -\sqrt n\mathbb E_n\varphi^{ep}_\pi
    \right|=o_p(1).
\]
The second display of the theorem follows from the Donsker limit of
$\{\varphi^{ep}_\pi:\pi\in\mathbf\Pi\}$ and the continuous mapping theorem.

It remains to compare the two Gaussian bounds. Let
$\psi_\pi=\varphi^{tp}_\pi-\varphi^{ep}_\pi$. A direct calculation gives
\[
    \psi_\pi(Z)
    =
    \big[m_1(X)\{\omega_1(X)T-1\}
    -m_0(X)\{\omega_0(X)(1-T)-1\}\big]\pi(X)
    +m_0(X)\{\omega_0(X)(1-T)-1\}.
\]
Thus $\psi_\pi$ is a bounded function of $(T,X)$ and
$\mathbb E[\psi_\pi\mid X]=0$. Also,
$\mathbb E[\psi_\pi\varphi^{ep}_{\pi'}]=0$ for every $\pi,\pi'$: the residual
part of $\varphi^{ep}_{\pi'}$ has conditional mean zero given $(T,X)$, and the
remaining part is a function of $X$.

The class $\{\psi_\pi:\pi\in\mathbf\Pi\}$ is again VC-type and hence
pre-Gaussian. Let $U^{tp}$ be a centered Gaussian process with covariance
$\mathbb E[\psi_\pi\psi_{\pi'}]$, independent of $G_{ep}$. The zero
cross-covariance just shown implies
\[
    G_{tp}\overset{d}=G_{ep}+U^{tp}.
\]
Finally, conditional Jensen's inequality gives, for each $\pi$,
\[
    \mathbb E\left[|G_{ep}(\pi)+U^{tp}(\pi)|\mid G_{ep}\right]
    \ge |G_{ep}(\pi)|.
\]
Using the usual separable version of the processes and then taking the
supremum over the countable determining subclass of $\mathbf\Pi$,
\[
    \mathbb E\left[\sup_{\pi\in\mathbf\Pi}|G_{tp}(\pi)|\right]
    =
    \mathbb E\left[\sup_{\pi\in\mathbf\Pi}|G_{ep}(\pi)+U^{tp}(\pi)|\right]
    \ge
    \mathbb E\left[\sup_{\pi\in\mathbf\Pi}|G_{ep}(\pi)|\right].
\]
This completes the proof.
\end{proof}

\section{Additional empirical results}
\label{app:empirics}

This section reports additional results for the commitment-savings
application: a comparison of the estimated rules
across the three criteria under the specification in the main text, and welfare--divergence frontiers as well as reassignment patterns for two alternative assignment specifications.

\begin{figure}[htbp!]
\centering
\includegraphics[width=0.78\linewidth]{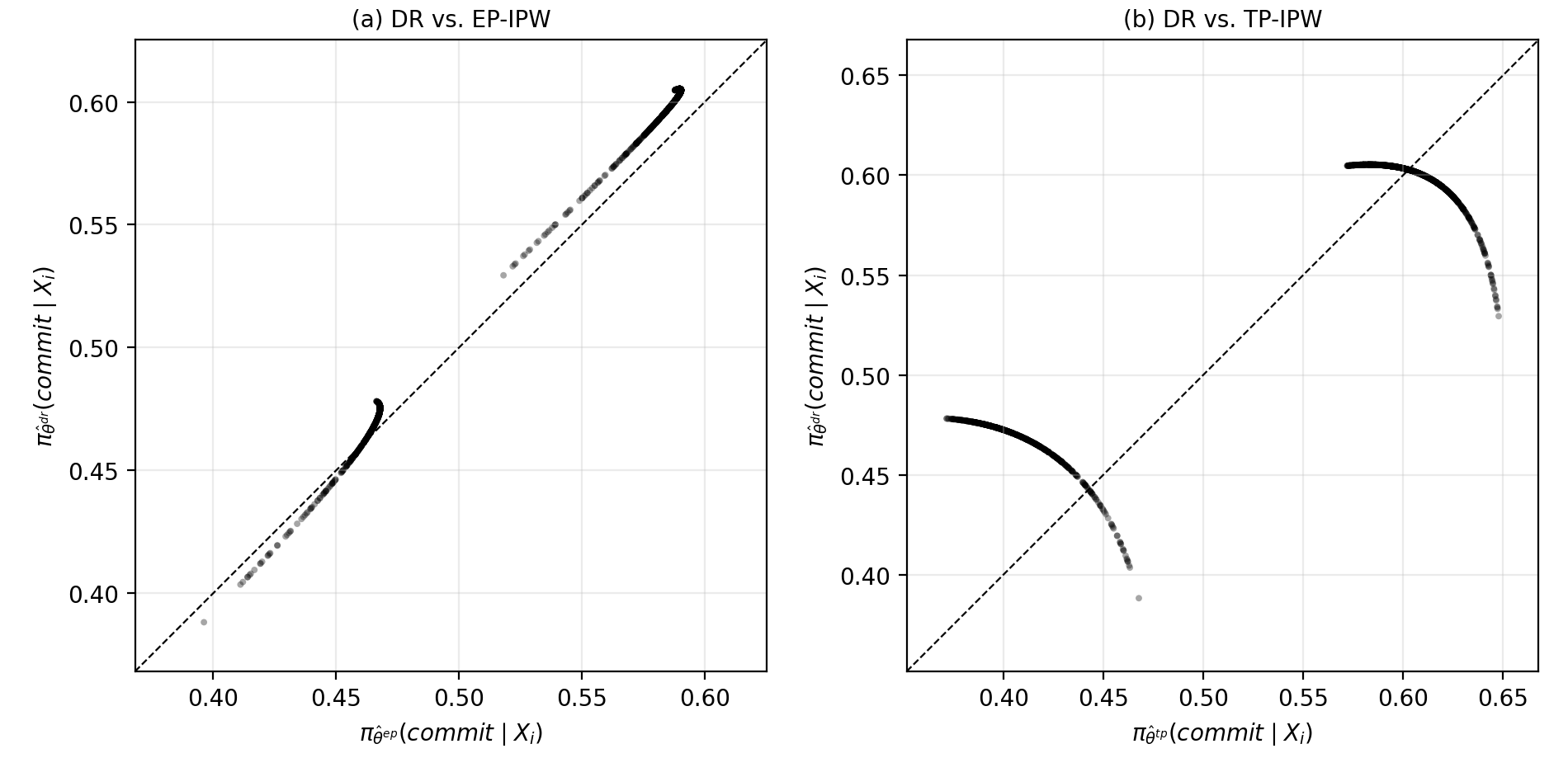}
\caption{Pairwise commitment-probability comparisons}
\caption*{At \(c=1.0\), primary
specification. Each point is one individual; the dashed line is the
45-degree line. Since active takes only two values, the population splits cleanly: the lower-left cluster is the clients with active = 1, and upper-right cluster is the clients with active = 0.}
\label{fig:aky_rule_similarity}
\end{figure}

\paragraph{Rule similarity across estimators.}\label{app:empirical-rule-similarity}
The estimators differ in finite samples not only in the value they assign to
a rule but in the rule itself. Figure~\ref{fig:aky_rule_similarity} compares
the per-individual commitment probabilities for each pair of estimators on
the primary specification at \(c=1.0\), and
Table~\ref{tab:aky_rule_similarity_summary} summarizes the absolute gaps.

\begin{table}[htbp!]
\centering
\renewcommand{\arraystretch}{1.20}
\begin{tabular}{lcccc}
\toprule
comparison & mean \(|{\Delta}|\) & median \(|{\Delta}|\) &
             p95 \(|{\Delta}|\) & share \(|{\Delta}| > 0.05\) \\
\midrule
DR vs EP-IPW       & 0.010 & 0.012 & 0.016 & 0.000 \\
DR vs TP-IPW       & 0.035 & 0.024 & 0.091 & 0.287 \\
EP-IPW vs TP-IPW   & 0.035 & 0.027 & 0.084 & 0.301 \\
\bottomrule
\end{tabular}
\caption{Absolute differences between the per-individual commitment
probabilities}
\label{tab:aky_rule_similarity_summary}
\end{table}

DR and EP-IPW produce almost identical rules: their commitment probabilities
differ by 0.010 on average, and no individual has a gap above 0.05. Both
differ from TP-IPW by a comparable margin, about 0.035 on average, with
roughly thirty percent of the sample above 0.05. This is the finite-sample
counterpart of the efficiency results in Section~\ref{sec:welfare_est}: the
two efficient criteria share the same first-order limit and select nearly the
same rule, while true-propensity weighting retains the treatment-assignment
noise component that the balancing weights remove, and that noise moves the
estimated rule.

\paragraph{Alternative assignment specifications.}
The main-text specification uses log income per capita and an indicator for
recent account activity. We rerun the full pipeline for two alternatives,
each pairing log income per capita with a second assignment covariate: the
number of household members, and an education indicator. As in
Section~\ref{subsec:choosing-lambda}, \(c\) is a preference parameter, and we
report the full welfare--divergence frontier for each specification.

\begin{figure}[htbp!]
\centering
\includegraphics[width=\linewidth]{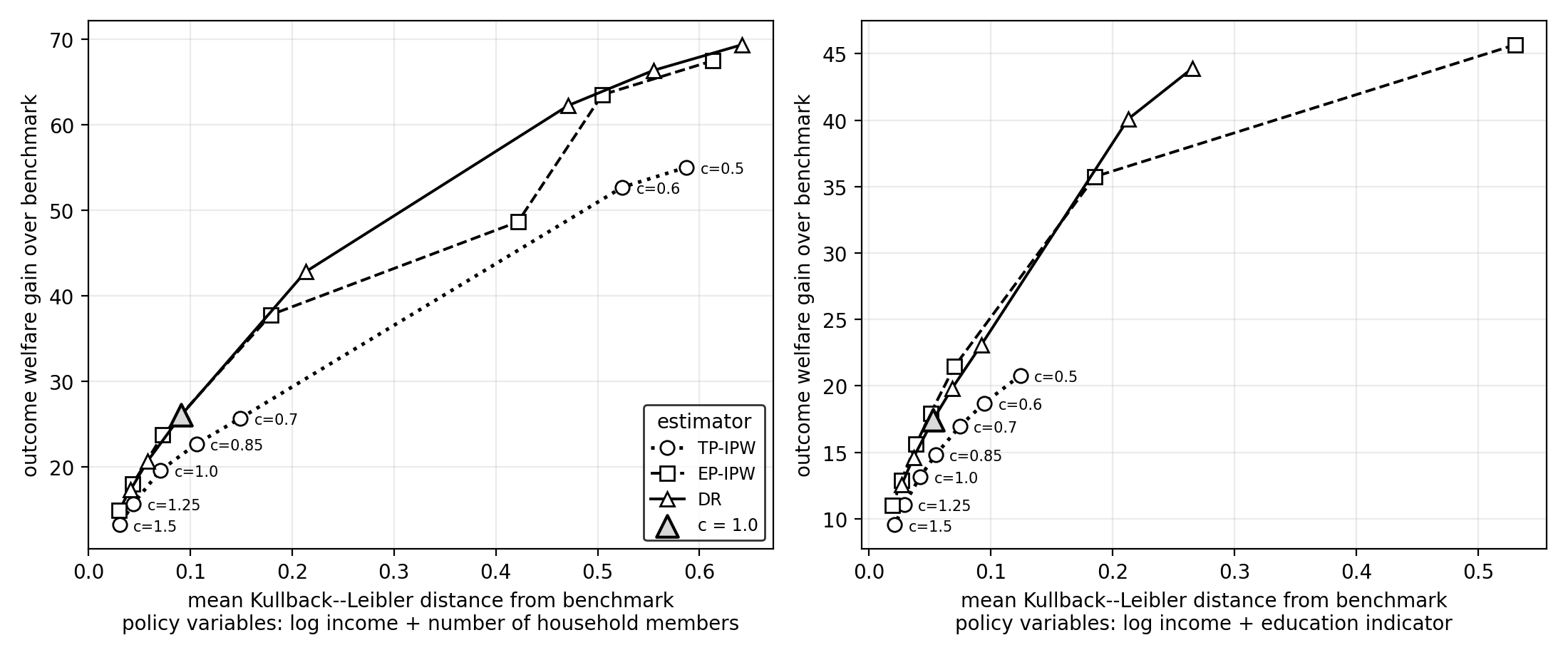}
\caption{Welfare--divergence frontiers for the two alternative
specifications}
\caption*{Each point is a benchmark-centered softmax rule estimated at
one value of \(c\); the vertical axis is the twelve-month savings gain over
the experimental rule, and the horizontal axis is the average KL divergence
from it. Labels on the TP-IPW curve indicate the value of \(c\); the shaded
triangle on the DR curve marks \(c=1.0\).}
\label{fig:aky_appendix_frontiers}
\end{figure}

Figure~\ref{fig:aky_appendix_frontiers} reports the frontiers. Both
specifications reproduce the qualitative ordering of the main specification:
EP-IPW lies above and slightly to the left of TP-IPW, and the DR frontier
closely tracks EP-IPW. The income-plus-household-size specification supports
a larger estimated outcome gain at every \(c\) than the main specification;
the income-plus-education specification is intermediate.

\begin{figure}[htbp!]
\centering
\includegraphics[width=\linewidth]{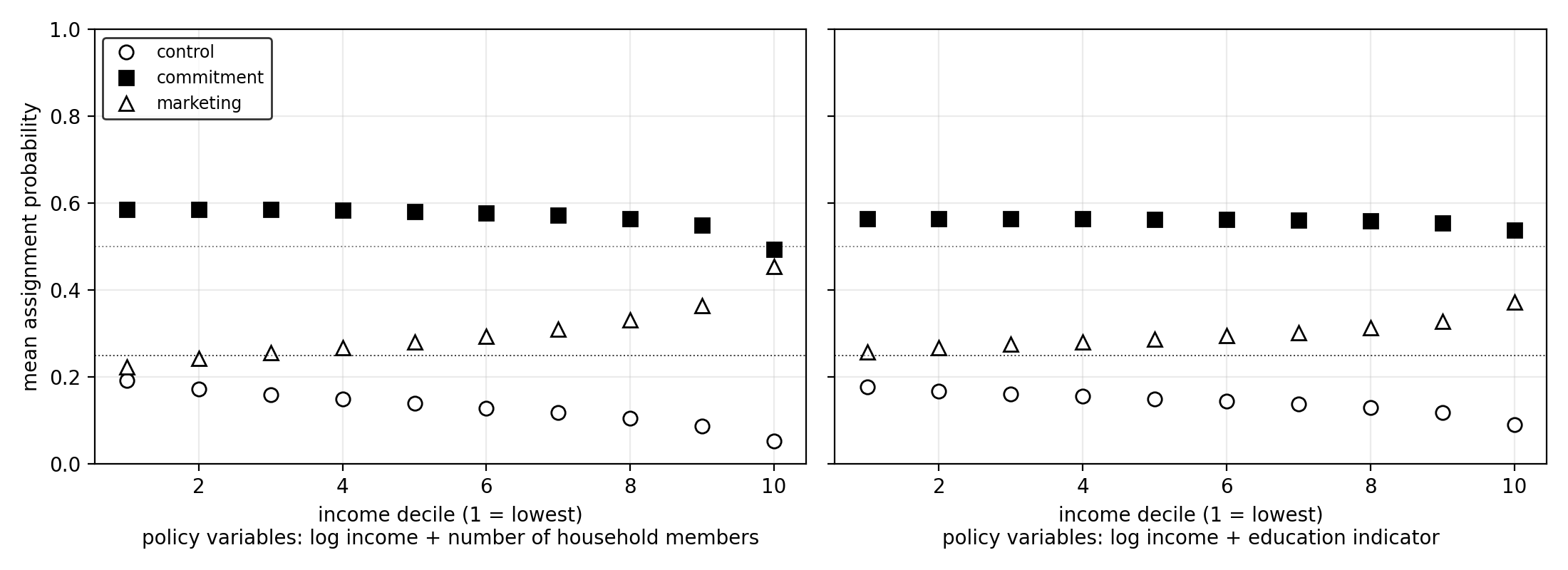}
\caption{DR rule assignment probabilities by income decile}
\caption*{At \(c=1.0\) for
the two alternative specifications, with the second assignment covariate
fixed at its sample median. Dotted reference lines mark the benchmark
probabilities \(\pi^{\mathrm b}=(0.25,\,0.50,\,0.25)\).}
\label{fig:aky_appendix_probs_by_income}
\end{figure}

Figure~\ref{fig:aky_appendix_probs_by_income} reports the DR rule's
assignment probabilities by income decile under each alternative
specification, with the second assignment covariate fixed at its sample
median (five household members in the first panel; education indicator equal
to one in the second). The pattern matches the main-text figure: the
commitment probability stays close to the benchmark value \(0.50\) across
deciles, while the control probability falls and the marketing probability
rises with income.

\section{Auxiliary lemmas} \label{sec:aux}

We use the following standard covering bound for VC classes.
\begin{lemma}[Theorem 2.6.7 of \citealp{van1996weak}]
\label{lm:covering-number-VC}
For a class of functions $\mathcal F$ with measurable envelope $F$ and finite
VC dimension $\operatorname{VC}(\mathcal F)$, there is a universal constant
$C$ such that, for every probability measure $Q$ with $\|F\|_{Q,2}>0$,
\[
    N\left(\epsilon\|F\|_{Q,2},\mathcal F,L_2(Q)\right)
    \le
    (C/\epsilon)^{2\operatorname{VC}(\mathcal F)},
    \qquad
    0<\epsilon<1.
\]
\end{lemma}

\end{document}

%% file: preamble.tex
\usepackage[english]{babel}
\usepackage[utf8]{inputenc}
\usepackage[authoryear,longnamesfirst]{natbib}
\usepackage[margin=1in]{geometry}
\usepackage{setspace}

\usepackage{amsmath}
\allowdisplaybreaks
\usepackage{amsfonts}
\usepackage{amsthm}
\usepackage{mathrsfs}
\usepackage{float}
\usepackage{amstext}
\usepackage{comment}
\usepackage{epsfig}
\usepackage{latexsym}
\usepackage{epstopdf}
\usepackage{bm}
\usepackage{algorithm}
\usepackage{algpseudocode}
\usepackage{amssymb}
\usepackage{url}
\usepackage{upgreek}
\usepackage{enumitem}

\usepackage{multirow}
\usepackage{multicol}
\usepackage{babel}
\usepackage{array}
\usepackage{rotating}
\usepackage{longtable}
\usepackage{float}
\usepackage{booktabs}
\usepackage{lscape}
\usepackage{graphicx} 
\usepackage{subcaption}

\usepackage{xr-hyper}
\usepackage[colorlinks=true,linkcolor=blue,citecolor=blue,urlcolor=blue]{hyperref}

\newtheorem{theorem}{Theorem}
\newtheorem{remark}{Remark}
\newtheorem{assumption}{Assumption}
\newtheorem{lemma}{Lemma}

\newtheorem{proposition}{Proposition}

\DeclareMathOperator{\tr}{tr}

\DeclareMathOperator{\Var}{Var}
\DeclareFontFamily{U}{mathx}{}
\DeclareFontShape{U}{mathx}{m}{n}{<-> mathx10}{}
\DeclareSymbolFont{mathx}{U}{mathx}{m}{n}
\DeclareMathAccent{\widecheck}{0}{mathx}{"71}